\documentclass[a4paper,fleqn,usenatbib]{mnras}

\usepackage{newtxtext,newtxmath}

\usepackage[T1]{fontenc}
\usepackage{ae,aecompl}

\usepackage{graphicx}	
\usepackage{amsmath}	

\usepackage{verbatim}
\usepackage[utf8]{inputenc}
\usepackage{hyperref}
\usepackage[dvipsnames]{xcolor}
\usepackage{units}
\usepackage{textcomp}
\usepackage{colortbl}

\definecolor{Berry}{HTML}{FF2052}

\usepackage{pgfplots}
\usepackage{filecontents}

\allowdisplaybreaks

\title[Dust survival for the Cas~A reverse shock: Magnetic fields]{Dust survival rates in clumps passing through the Cas~A reverse shock -- II. The impact of magnetic fields}

\author[F. Kirchschlager et al.]{Florian Kirchschlager\href{https://orcid.org/0000-0002-3036-0184}{\includegraphics[trim=0cm -35cm 0cm 0cm, clip=true,width=0.27cm]{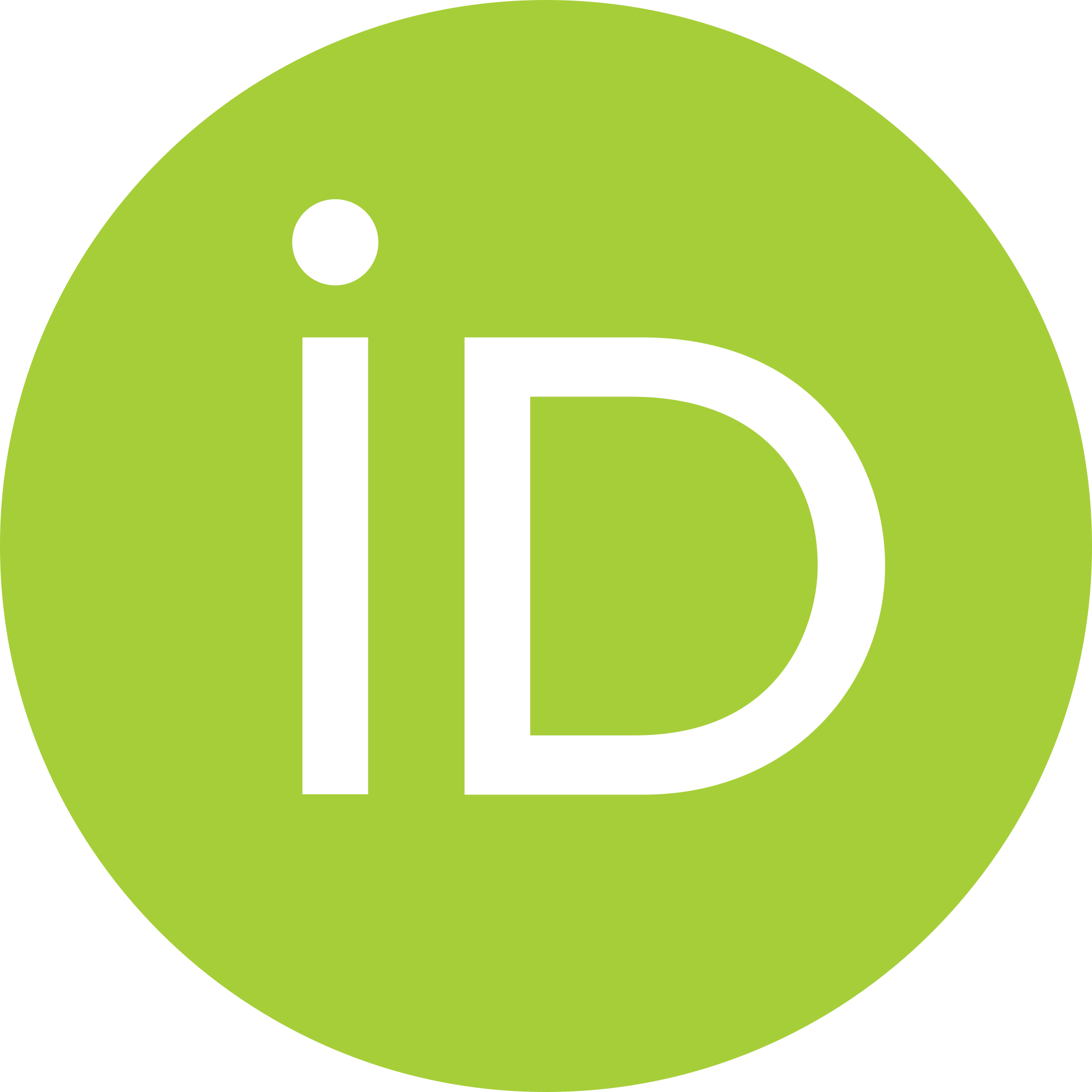}},$^{1,2}$\thanks{E-mail: florian.kirchschlager@ugent.be}
Franziska D. Schmidt,$^{2}$ 
M. J. Barlow\href{http://orcid.org/0000-0002-3875-1171}{\includegraphics[trim=0cm -35cm 0cm 0cm, clip=true,width=0.27cm]{Pics/orcidID.jpg}},$^{2}$
Ilse De Looze\href{http://orcid.org/0000-0001-9419-6355}{\includegraphics[trim=0cm -35cm 0cm 0cm, clip=true,width=0.27cm]{Pics/orcidID.jpg}},$^{1}$\newauthor 
Nina S. Sartorio\href{http://orcid.org/0000-0003-2138-5192}{\includegraphics[trim=0cm -35cm 0cm 0cm, clip=true,width=0.27cm]{Pics/orcidID.jpg}},$^{1}$\\
$^{1}$Sterrenkundig Observatorium, Ghent University, Krijgslaan 281-S9, B9000 Ghent, Belgium\\
$^{2}$Department of Physics and Astronomy, University College London, Gower Street, London WC1E 6BT, United Kingdom
}

\date{Accepted 2023 January 23. Received 2022 December 23; in original form 2022 October 14
}

\pubyear{2023}

\begin{document}
\label{firstpage}
\pagerange{\pageref{firstpage}--\pageref{lastpage}}
\maketitle

 \begin{abstract}
Dust grains form in the clumpy ejecta of core-collapse supernovae where they are subject to the reverse shock, which is able to disrupt the clumps and destroy the grains. Important dust destruction processes
include thermal and kinetic sputtering as well as fragmentation and grain vaporization. In the present study, we focus on the effect of magnetic fields on the destruction processes. We have performed magneto-hydrodynamical simulations using \textsc{\mbox{AstroBEAR}} to model a shock wave interacting with an ejecta clump. The dust transport and destruction fractions are computed using our post-processing code \textsc{\mbox{Paperboats}} in which the acceleration of grains due to the magnetic field and a procedure that allows partial grain vaporization have been newly implemented.
For the oxygen-rich supernova remnant Cassiopeia~A we found a significantly lower dust survival rate when magnetic fields are aligned perpendicular to the shock direction compared to the non-magnetic case. For a parallel field alignment, the destruction is also enhanced but at a lower level. The survival fractions depend sensitively on the gas density contrast between the clump and the ambient medium and on the grain sizes. For a low-density contrast of $100$, e.g., $5\,$nm silicate grains are completely destroyed while the survival fraction of $1\,\mu$m grains is $86\,$per~cent. For a high-density contrast of $1000$, $95\,$per~cent of the $5\,$nm grains survive while the survival fraction of $1\,\mu$m grains is $26\,$per~cent. Alternative clump sizes or dust materials (carbon) have non-negligible effects on the survival rate but have a lower impact compared to density contrast, magnetic field strength, and grain size.
 \end{abstract}

\begin{keywords}
supernovae: general -- ISM: supernova remnants -- dust, extinction -- magneto-hydrodynamics -- shock waves -- supernovae: individual: Cassiopeia A
\end{keywords}
 



\defcitealias{Kirchschlager2019b}{Paper~I}

\section{Introduction}
\label{100}
Observations have proven the formation of dust grains in the expanding remnants of core-collapse supernovae (SNe; e.g.,~\citealt{Lucy1989,Barlow2010, Gall2011, Matsuura2011, Matsuura2022, Gomez2012, Wesson2015,Bevan2017, DeLooze2017, NiculescuDuvaz2022}). The reverberation of that explosion is the reverse shock that passes through the ejecta remnant and which is energetic enough to potentially destroy large amounts of the freshly produced dust grains (e.g.,~\citealt{Nozawa2007, Bianchi2007, Nath2008, Silvia2010, Silvia2012, Biscaro2016, Micelotta2016, Martinez2018, Kirchschlager2019b, Slavin2020, Priestley2022}). The net amount of dust grains present is even more negatively affected given that the forward shock potentially destroys pre-existing dust in the circumstellar and interstellar medium (ISM; e.g.,~\citealt{Nozawa2006, Bocchio2014, Slavin2015,Dopita2016, Martinez2019, Priestley2021}). It is therefore still an open question whether SNe are a net dust producer or destroyer (\citealt{Kirchschlager2022}).

When the shock wave hits the dust grains embedded in the over-dense clumps of gas, the dust grain surfaces are bombarded by gas ions which causes various effects. Due to the conservation of momentum, the grains are accelerated in the shock direction, roughly anti-proportional to their size. At the same time, grain atoms can be ejected from the grain surfaces and the size dependent grain acceleration can lead to catastrophic collisions of grains of different sizes. These processes, sputtering and grain-grain collisions, can cause a substantial destruction of the dust grains and depend beside the dust properties also on the conditions of the surrounding gas. Moreover, dust grains in the supernova remnant (SNR) are electrically charged by the impacts of plasma particles (ions and electrons).
 
 The first paper of this series (\citealt{Kirchschlager2019b}, hereafter \citetalias{Kirchschlager2019b}) focused on the dust destruction by the reverse shock in Cassiopeia~A (Cas~A) for a range of clump densities. For this purpose, we developed the post-processing code \textsc{Paperboats} which computes the dust transport and dust destruction in a moving gas on the basis of the output of a (magneto-)hydrodynamical (MHD) code. Cas~A is a Galactic dusty SNR that has been studied extensively because of its relatively close distance (e.g.,~\citealt{Dwek1987,Gotthelf2001,Fesen2006, Rho2008, Arendt2014, Priestley2019a}). In the present paper, we concentrate on the influence of magnetic fields on the dust transport and destruction processes in Cas~A. When the shock hits the dusty clumps, charged grains start to spiral around the magnetic field lines (e.g.,~\citealt{Northrop1984}) and are thus deflected from their original motion caused by the streaming gas. Larger grain velocities and thus a higher risk of grain-grain collisions are expected. When the grains start to react to the presence of the magnetic fields, the relative velocities of gas and dust can increase.
 
SNRs older than $2000\,$yr tend to have magnetic fields that are tangentially oriented (parallel to the shock fronts in a spherical SNR), whereas young SNRs have radially oriented magnetic fields (\citealt{Dickel1976}). This agrees with the observed radial orientation in Cas~A (e.g.~\citealt{Rosenberg1970, Vink2022b}). It may well be that the fields closer to the shock fronts are tangentially oriented (\citealt{Jun1996, Bykov2020}), since the shock compresses and thus enhances the magnetic field component perpendicular to the shock direction. 

For Cas~A different observations have attempted to derive the magnetic field characteristics in the shocked or unshocked regions of the ejecta. 
An early estimate of the overall magnetic-field strength was based on the minimum energy argument and amounts to ${\sim}500\,\mu$G (\citealt{Rosenberg1970}) which is far above the common value for the warm diffuse phase of the ISM of $3\,\mu$G (see \citealt{Jones1996}). The observed widths of the \mbox{X-ray} synchrotron filaments of ${\sim}10^{ 17}\,$cm suggest that the local, downstream magnetic-field strengths are  $250-550\,\mu$G (\citealt{Vink2003, Berezhko2004, Bamba2005, Ballet2006, Helder2012}).
Arguments to explain the radio luminosity require magnetic fields of ${\sim}400-2000\,\mu$G (\citealt{Longair1994, Wright1999}). Alternative analyses based on radio, infrared, X-ray, and gamma-ray data suggest magnetic field strengths of $50-300\,\mu$G (\citealt{Araya2010}), $230-510\,\mu$G (\citealt{Saha2014}), and $200-400\,\mu$G (\citealt{Zirakashvili2014}). 
\cite{Kilpatrick2016} used  near-infrared multi-epoch data and derived magnetic field strengths in dense knots in the post-shock region of $1300-5800\,\mu$G. Recently, \cite{Domcek2021} estimated the magnetic-field strength to be below $\unit[1000]{\mu G}$ based on a break of the spectral slope in the near- to mid-infrared regime.

The differences of 1-3 orders of magnitude between the magnetic field strength derived from observations of Cas~A and in the ISM can be explained by either a larger magnetic field strength around Cas~A, e.g. as the result of the stellar wind of the progenitor (\citealt{Biermann1993}), or a rapid enhancement near the shock front due to gas compression. Moreover, the magnetic field strengths derived in the observational studies have been averaged over a certain area. On small scales, compression and turbulent motion can lead to significant magnetic field amplification and thus to even higher magnetic field strengths than observed.  In our study, we assume a moderate magnetic field strength of a few $\mu$G in the unshocked ejecta regions which will then be amplified to several 100 to a few $1000\,\mu$G in the post-shock gas. 
 
 The paper is organized as follows: In Section~\ref{600} we briefly discuss previous studies of magnetic fields in dust destruction simulations. In Section~\ref{300} the setup for the MHD simulations of the reverse shock impacting an over-dense clump of gas in the ejecta is presented. Section~\ref{400} describes the physics and properties which are important for the dust processing in the shocked clump, in particular the effect magnetic fields have on the dust dynamics and on the destruction efficiency. The results of our simulations with or without magnetic fields, the disruption of the clump and the destruction of the dust, are discussed in Section~\ref{450}. After giving a schematic overview about which dust grain sizes are destroyed at various clump densities or magnetic field strengths and a discussion about the importance of grain-grain-collisions and about the destruction in the entire remnant (Section~\ref{500}), we conclude with a summary of our findings in Section~\ref{700}.


 \section{Previous studies}   
 \label{600} 
 In recent years a number of studies have investigated the dust destruction by the passage of the reverse shock in SNRs or by the forward shock in the ISM. We attempt to give a brief overview of the implementation of magnetic fields and their impact on charged dust grains in these studies.

In general, magneto-hydrodynamic turbulence can trigger grain shattering (\citealt{Hirashita2010}), producing an excess of small grains that can be rapidly sputtered in the shocked hot gas. Moreover, charged particles moving in a magnetic field experience additional acceleration (betatron acceleration, Lorentz force) which critically modifies the grain dynamics (\citealt{Jones1994,Jones1996,Guillet2007,Slavin2004, Bocchio2014}). The gyration around the magnetic fields tends to strengthen the gas-to-dust coupling which prevents the ejection of large grains from the SNRs into the ISM. Simultaneously, the additional acceleration increases the local relative velocities between dust grains and gas and can thus enhance the rates of grain-grain collisions and sputtering (\citealt{Shull1978}).

The modified grain dynamics makes it important to trace the trajectories in the SNRs and in the ISM. To study the dust processing in the ISM shocked by SN blast waves,  \cite{Slavin2015} and \cite{Hu2019} conducted hydrodynamical simulations. The magnetic field is included in both studies in an approximate way. In the 1D model of \cite{Slavin2015} the magnetic pressure term is proportional to the density which is applicable to the perpendicular component of the magnetic field. At the same time, this approach ignores magnetic tension. They adopted a uniform ISM magnetic field strength of $3\,\mu$G and predicted a reduced dust destruction due to a lower gas compression if lower magnetic fields are taken into account. In the 3D models of \cite{Hu2019}, magnetic fields are not considered, however, betatron acceleration is realized under the assumption of flux-freezing and a strong planar shock, which again allows only for perpendicular  magnetic fields and a proportionality between gas density and magnetic field strength. In their study, they found an increased ISM dust destruction rate when betratron acceleration is considered which is caused by increased sputtering rates in the compressed gas.
 
 
 \cite{Fry2020} investigated the dust injection from a SNR into its environment. They assumed turbulent magnetic fields in the interstellar medium (ISM) which are amplified by the SNR shock, while the SN wind and ejecta fields are negligible. They found that magnetic fields are crucial for the dynamical description of the ejecta grains. Charged grains formed in the SNR can decouple from the gas and are prevented from traversing the contact discontinuity which separates the shocked ejecta from the shocked ISM. Instead, these grains are reflected and trapped within the ejecta, limiting the SN dust injection into the circumstellar ISM.
 
 \cite{Micelotta2016} studied the dust destruction by the reverse shock in Cas~A and adopted an initial magnetic field of $1\,\mu$G perpendicular to the shock direction in the unshocked ejecta and clumps. This is in agreement with \cite{Sutherland1995} who assumed this strength for the Cas~A ejecta clumps.   Although such a magnetic field evokes betatron acceleration of the charged grains, they did not consider magnetic forces acting on the dust grains. \cite{Bocchio2016} argued that betatron acceleration is probably not relevant for dust produced in the SNRs: polarimetry observations (e.g.,~\citealt{Dunne2009}; see also references in Section~\ref{100}) and numerical simulations (e.g.,~\citealt{Inoue2013, Schure2014}) show hints of a radial alignment of the magnetic field inside the ejecta. If the main motion of gas and dust is also mostly radially oriented, betatron acceleration is not effective. In contrast,  \cite{Bocchio2014} considered betatron acceleration and assumed a magnetic field strength of $B_0=3\,\mu$G to study the dust destruction  in the warm ionised ISM, however, they did not report on the influence of the magnetic field.
 
 \cite{Martinez2018} neglected the impact of magnetic fields on the dust grain motions as the grains tend to be neutral at  high gas temperatures prevailing in the shocked regions  (${>}2 \times 10^5\,$K for silicate/graphite grains; \citealt{McKee1987}). However, we note that the efficiency of grain charging significantly depends on the shock velocity, gas density and gas composition (see, e.g., \citealt{Fry2020}, \citetalias{Kirchschlager2019b}). \cite{Slavin2020} ignored magnetic ﬁelds in their simulations of dust grains in the ejecta of SNRs due to the uncertainties in the charging as well as in the magnetic field strengths and morphology. However, they note that magnetic fields could reduce the escape of dust grains from the ejecta under particular conditions, similar to the findings of \cite{Fry2020}.


The present study is the first work to consider magnetic fields and magnetic field forces acting on the dust dynamics and dust destruction processes in the clumpy ejecta of a SNR. In particular, the impact on the dust transport, sputtering and grain-grain collisions is investigated when the dusty clump is disrupted and processed by the reverse shock in Cas~A.

\section{Magneto-hydrodynamical simulations}
\label{300}
To simulate the dynamical evolution of a reverse shock impacting a clump of ejecta material in the SNR, the MHD code \textsc{\mbox{AstroBEAR}}\footnote{\href{https://www.pas.rochester.edu/astrobear/}{https://www.pas.rochester.edu/astrobear/}} (\citealt{Cunningham2009, Carroll-Nellenback2013}) was used, a highly parallelized, multidimensional adaptive mesh refinement code which solves the conservative equations of magneto-hydrodynamics on a Cartesian grid (see e.g. \citealt{Poludnenko2002, Cunningham2009, Kaminski2014, Fogerty2016,Fogerty2017, Fogerty2019}). \textsc{\mbox{AstroBEAR}} models only the gas-phase of the ejecta environment, for the analysis of the dust evolution we will employ the post-processing code \textsc{\mbox{Paperboats}} (\citetalias{Kirchschlager2019b}; Section~\ref{400}).

\subsection{Model setup}
 \label{sec_mod_setup}
The  temporal and spatial evolution of the dust is highly affected by the local gas density distribution. To investigate the destruction of the clumps at sufficiently high resolution, we do not model the entire three-dimensional remnant but a section of it, in which only one clump is impacted by the reverse shock. This scenario is called the cloud-crushing problem (\citealt{Woodward1976}) and was already applied by \cite{Silvia2010,Silvia2012} and \cite{Kirchschlager2019b,Kirchschlager2020} to study the dust survival in SNRs. The impact of the reverse shock on a single clump is assumed to happen for all the clumps existing in the ejecta so that results of the cloud-crushing problem can be applied and extrapolated to the entire remnant.
 
\subsubsection{Pre-shock conditions}
\label{2001}
 In the cloud-crushing problem a planar shock is driven into an over-dense clump of gas which is embedded in a low-density gaseous medium (Fig.~\ref{sketch_clump}). We adopt parameters that represent the reverse shock and the clumpy ejecta in Cas~A. In the pre-shock gas, the ambient medium has a number density $n_\text{am}=\unit[1]{cm^{-3}}$ of gas particles (oxygen) and a temperature $T_\text{am}=\unit[10^4]{K}$. The embedded clump has a spherical shape with radius $R_\text{cl} = \unit[10^{16}]{cm}\approx\unit[668.5]{au}$ (except for Section~\ref{sec_cs} where we doubled the clump size), a temperature of $T_\text{cl}=\unit[10^2]{K}$, and a uniform gas number density of $n_\text{cl}=\chi n_\text{am}$. The initial gas density contrast $\chi=n_\text{cl}/n_\text{am}$ is crucial and dominates the total simulation time and the size of the computational domain (see below), but also the dust survival rate (see Section~\ref{511}). We vary the density contrast $\chi$ between 10 and 1000. The pre-shock magnetic field $B_0$ and its orientation are the same in the ambient medium and in the clump. We vary the magnetic field strength between 0 and $\unit[10]{\mu G}$ and the orientation is perpendicular to the shock direction (except for Section~\ref{magn_para} where we study a parallel alignment of the magnetic field). The shock velocity in the ambient medium is adopted to be $v_\text{sh} = \unit[1600]{km/s}$ following the analytical result of \cite{Micelotta2016}. In the ambient medium it is fixed for each simulation, independent of the density contrast and the initial magnetic field strength, while it is decelerated in the over-dense clump to $\sim\!\chi^{-0.5} v_\text{sh}$. The mean molecular weight of the pre-shock gas is set to \mbox{$\mu=16.0$,} corresponding to a pure oxygen gas. All pre-shock parameters are listed in Table~\ref{tab_1}.

\subsubsection{Simulation time} 
Besides physical parameters we also have to set numerical parameters like the simulation time and the computational domain. 

\begin{figure}
\centering
\includegraphics[trim=0.3cm 0cm 0cm 0cm, clip=true,page=1,width=1.0\linewidth]{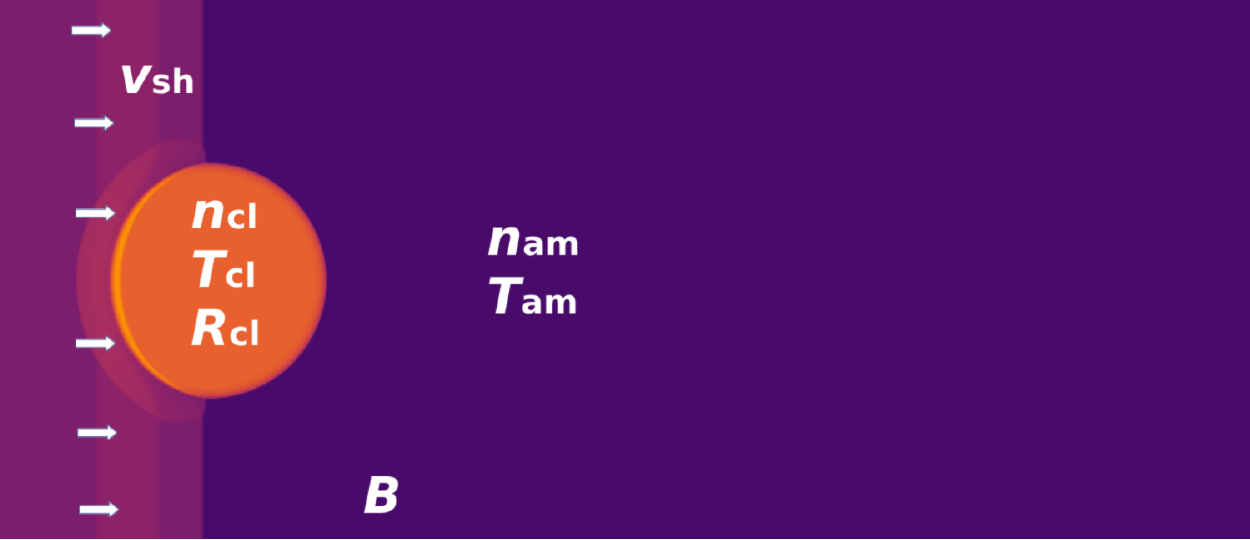}
\caption{Sketch of the cloud-crushing problem: reverse shock (magenta) with shock velocity $v_\text{sh}$, impacting on an over-dense clump of gas (orange; $n_\text{cl},T_\text{cl},R_\text{cl}$) embedded in a low-density gaseous medium (purple; $n_\text{am},T_\text{am}$). The magnetic field strength in the gas is $B$.}
\label{sketch_clump} 
 \end{figure}

At the beginning of the simulation ($t=0$), the clump is at rest, embedded in the ambient medium, and the centre is placed at a distance of $1.5\,R_\text{cl}$ in front of the shock front to prevent material swept up by the bow shock from leaving the domain in the direction contrary to the shock propagation (Fig.~\ref{sketch_clump}). We follow the evolution of the shocked clump for $3\,\tau_\text{cc}$ after the first contact of the shock with the clump, where
 \begin{align}
 \tau_\text{cc}=\chi^{0.5}R_\text{cl}/v_\text{sh} \label{cloudcrushingtime}                                                                                                                                                                                                                                                                                                                                                                                                                                                                                                                                                                                                                                                                                                                                                                                                                                                                                                                                                                                                                                                                                                                                \end{align}
 is the cloud-crushing time as defined by \cite{Klein1994} which gives the characteristic time for the shock to penetrate the clump. $3\,\tau_\text{cc}$ is a commonly used value to investigate post-shock structures during which the clump is totally disrupted by the shock. 
%
%
%
%
%
%
%
Taking into account some extra time the shock needs to reach the clump, we set the total simulation time to $t_\text{sim}=(3\chi^{0.5}+1)R_\text{cl}/v_\text{sh}$ which is slightly more than three cloud-crushing times. The simulation time for the density contrast $\chi=100$ is then $\unit[61.5]{yr}$ which is roughly $\unit[20]{\%}$ of the total age of Cas~A (${\sim}340-350\,$yr; \citealt{Fesen2006}). For comparison, the simulation time for $\chi=10$ and 1000 are $20.8\,$yr and $190.2\,$yr, respectively.
 
 \begin{table}
 \centering
 \caption{Initial conditions for the cloud-crushing simulations carried out in this study. The parameters in brackets are considered in Sections~\ref{sec_cs} and \ref{magn_para} only.}
 \begin{tabular}{ l l}
 \hline\hline
 Parameter&Values\\\hline
Shock velocity $v_\text{sh}$& $\unit[1600]{km\,s^{-1}}$ \\
Density contrast $\chi = n_\text{cl}/n_\text{am}$ &10, 20, 50, 100, 200, 300, 500, 1000\\  
Clump gas temperature $T_\text{cl}$& $\unit[10^2]{K}$\\
Clump radius $R_\text{cl}$ &$\unit[10^{16}]{cm}$, ($\unit[2\times10^{16}]{cm})$ \\
Amb. med. gas density  $n_\text{am}$ & $\unit[1]{cm^{-3}}$\\  
Amb. med. gas temperature $T_\text{am}$\hspace*{-0.3cm}  & $\unit[10^4]{K}$\\  
 Magnetic field strength $B_0$     & $\unit[0, 1, 3, 5, 10]{\mu G}$\\  
 Magnetic field orientation   & perpendicular, (parallel)\\
 Mean molecular weight $\mu$   & $16.0$\\\hline
 \end{tabular}
 \label{tab_1}
 \end{table}
 
\subsubsection{Computational domain}
 We consider 2D MHD simulations due to the large computational effort for highly resolved 3D post-processing simulations (see Section~\ref{400} for the dimensionality of the post-processing). In order to ensure that the clump material does not flow out of the domain at the back end during the simulation time $t_\text{sim}$, the length and the width of the domain are set to $l_\text{box}=15\,R_\text{cl} = \unit[0.049]{pc}$  and $w_\text{box}= 6\,R_\text{cl} = \unit[0.019]{pc}$ for nearly all density contrasts $\chi$. However, for the highest density contrast $\chi=1000$ we set $l_\text{box}=21\,R_\text{cl} = \unit[0.068]{pc}$ and $w_\text{box}= 7\,R_\text{cl} = \unit[0.023]{pc}$, as the simulation time is significantly longer. The computational domain for all density contrasts consists of $1500\times600$ cells\footnote{Only for $\chi=1000$, we adopt a larger grid consisting of $2100\times700$ cells to enable the same resolution of 100 cells per clump radius as for the other density contrasts. } such that there are 100 cells per clump radius. This is a significant improvement compared to \citetalias{Kirchschlager2019b} in which the resolution was 20 cells per clump radius. The physical resolution is now $\Delta_\text{cell}=\unit[10^{14}]{cm}$ ($\sim\unit[6.7]{au}$) per cell.  Outflow boundary conditions are used on all sides of the domain, with the exception of the left boundary (Fig.~\ref{sketch_clump}) which used an inflow boundary for injecting a continuous post-shock wind into the domain.

\subsubsection{Gas cooling} 
The MHD simulations consider cooling of the oxygen-rich gas. We use the same cooling function as in \citetalias{Kirchschlager2019b} (see Figure~3 therein): a combination of the cooling curves of \cite{Sutherland1995} for temperatures below $10^4\,$K and the cooling derived using CHIANTI \citep{Delzanna2015} for a gas of pure oxygen in ionization equilibrium for the temperature range $T_\text{gas} = 10^4-10^9$K. The cooling at lower temperatures is dominated by line emission and at higher temperatures by collisional ionization, bremsstrahlung emission and contributions from radiative recombination \citep{Raymond2018}. As outlined by \cite{Silvia2010}, cooling reduces the gas pressure in the shocked clumps and facilitates the formation of cold dense nodules, which can both have an impact on the dust survival rate.

\subsection{Post-shock conditions}
The initial conditions of the cloud-crushing problem are given in Section~\ref{2001}. Based on these pre-shock values, the post-shock quantities are calculated by \mbox{AstroBEAR} using the Rankine-Hugoniot jump conditions (RHJ conditions). We had to implement the RHJ conditions in \mbox{AstroBEAR} for the MHD case of the cloud-crushing problem. The basic expressions are outlined in Appendix~\ref{app_RHJC}.

\section{Dust evolution}
\label{400}
We use our post-processing code \textsc{\mbox{Paperboats}} (\citetalias{Kirchschlager2019b})
to study the evolution of dust in a moving gas. Based on the temporally and spatially resolved gas density, temperature, velocity, and magnetic field provided by \textsc{\mbox{AstroBEAR}} in 2D, we investigate the dust transport and derive the dust destruction rate. Up until now \textsc{\mbox{Paperboats}} was able to treat only non-magnetic effects. For the present study we have updated and extended \textsc{\mbox{Paperboats}} to allow us to consider the Lorentz force on a charged dust grain in a magnetic field which affects the dust transport as well as dust destruction processes. 

It is important to consider grain-grain collisions in 3D as this will affect the grain cross-sections and collision probabilities. Moreover, the gyration of charged grains around the magnetic field lines require a 3D setup. Therefore, the 2D MHD simulations are extended here to 3D assuming a single cell in the $z$-direction while the gas velocity in $z$-direction is zero. We apply periodic boundary conditions in $z$-direction to avoid artificial loss of dust grains when grains cross and overcome the single cell.

The methodology of the code was first presented in \mbox{\citetalias{Kirchschlager2019b}}. We take up on this and describe here the implementation of magnetic fields in \textsc{\mbox{Paperboats}}. 
 

%

\subsection{Dust transport}
As outlined in \citetalias{Kirchschlager2019b} (Section~4.3) the dust velocity $\mathbf{v_\text{dust}}(t+\Delta t)$ at time $t+\Delta t$ is determined by
 \begin{align}
 \mathbf{v_\text{dust}}(t+\Delta t) = \mathbf{v_\text{dust}}(t) + \sum_{i=1}^{10} \mathbf{a}_\text{acc,total}\left(t'\right) \,\frac{\Delta t}{10},
 \end{align}
where $\mathbf{v_\text{dust}}(t)$ is the grain velocity at time $t$, $\Delta t$ is the time-step between two output frames of the hydrodynamical simulations, and $\mathbf{a}_\text{acc,total}\left(t'\right)$ is the total acceleration experienced at time $t'=t+\Delta t(i-1)/10$. We assume that the conditions of the surrounding gas are constant during $\Delta t$. For the sake of higher velocity accuracy, the time interval $\Delta t$ is divided into ten equally-sized intervals in which the acceleration is calculated. $\Delta t/10$ is then the smallest time interval. The total number of these $\Delta t/10$ time intervals is fixed to 1250 for the calculation of the dust dynamics for all our simulations. This number has been shown to be appropriate for the dust dynamics in a flowing, shocked gas (\citetalias{Kirchschlager2019b}). The total acceleration at time $t'$ of an individual dust grain of mass $m$ is made up of a drag term (non-magnetic) and a Lorentz term (magnetic),
 \begin{align}
\mathbf{a}_\text{acc,total}\left(t'\right) = \frac{\mathbf{F}_\text{drag}\left(t'\right)}{m} + \frac{\mathbf{F}_\text{Lorentz}\left(t'\right)}{m}.
\end{align}
The drag term is given by collisional drag and plasma drag (\citealt{Baines1965, Draine1979}) and we refer to \citetalias{Kirchschlager2019b} for full details. The Lorentz force\footnote{Equation~(\ref{Lorentz}) is in SI units; for cgs units the charge has to be divided by the speed of light $c$.} is defined as
\begin{align}
 \mathbf{F}_\text{Lorentz} =Q_\text{grain}\, \mathbf{v}_\text{rel}\times \mathbf{B}, \label{Lorentz}
\end{align}
where $Q_\text{grain}$ is the dust grain charge, $\mathbf{B}$ is the magnetic field, and $\mathbf{v}_\text{rel}$ is the relative velocity between the dust and the surrounding gas (and thus between the dust and the magnetic field which is coupled to the gas) at time $t'$,  $\mathbf{v}_\text{rel}\left(t'\right) = \mathbf{v}_\text{gas}\left(t'\right) - \mathbf{v}_\text{dust}\left(t'\right)$.
In order to determine the dust transport and the grain trajectories, we change the coordinate system for the sake of simplicity and multiply $\mathbf{v}_\text{rel}$ and $\mathbf{B}$ with a rotation matrix $\mathbf{M}$, respectively, so that $\mathbf{B}$ is directed in $z$-direction only, \mbox{$\mathbf{B}=(0,0,B_z)$}, and the component of the relative velocity perpendicular to the magnetic field is directed in $y$-direction only, $\mathbf{v}_\text{rel}=\left(0, v_y, v_z\right)$. The parts of  $\mathbf{v}_\text{rel}$ perpendicular and parallel to the magnetic field are then $v_\text{perp}=v_y$ and $v_\text{para}= v_z$, respectively. Considering the Lorentz force and using $\omega=Q_\text{grain}\,|\mathbf{B}|/m$ as angular velocity\footnote{Please note, $\omega$ can have a negative sign due to the grain charge.} and \mbox{$R_\text{gyro} = v_\text{perp}/|\omega|$} as the gyration radius (Larmor radius), the position of the dust grain at time $t'+\Delta t/10$ amounts to
\begin{align}
\mathbf{r}(t'+\Delta t/10) =
&\left(
 \begin{array}{c}
x(t'+\Delta t/10)\\
y(t'+\Delta t/10)\\
z(t'+\Delta t/10)
\end{array}
\right)
   \nonumber\\ =& 
\left(
  \begin{array}{l}
x(t') +  R_\text{gyro} \sin{(\omega\,\Delta t/10)}\\
y(t') +  R_\text{gyro} \cos{(\omega\,\Delta t/10 -\pi/2)}\\
z(t') +  v_\text{para}\,\Delta t/10
  \end{array}
  \right). \label{eq_change}
\end{align}
The motion of the dust grain projected onto the $x$-$y$-plane of that system is a circle while the motion in $z$-direction is linear. 
The direction of the dust transport between $t'$ and $t'+\Delta t/10$ is then given by a displacement vector,
\begin{align}
 \Delta \mathbf{r}= \mathbf{r}(t'+\Delta t/10) - \mathbf{r}(t') \label{eq_transport}.
\end{align}

   \begin{figure}
   \resizebox{\hsize}{!}{
  \includegraphics[trim=1.0cm 7.1cm 7.5cm 6.7cm, clip=true, page=6]{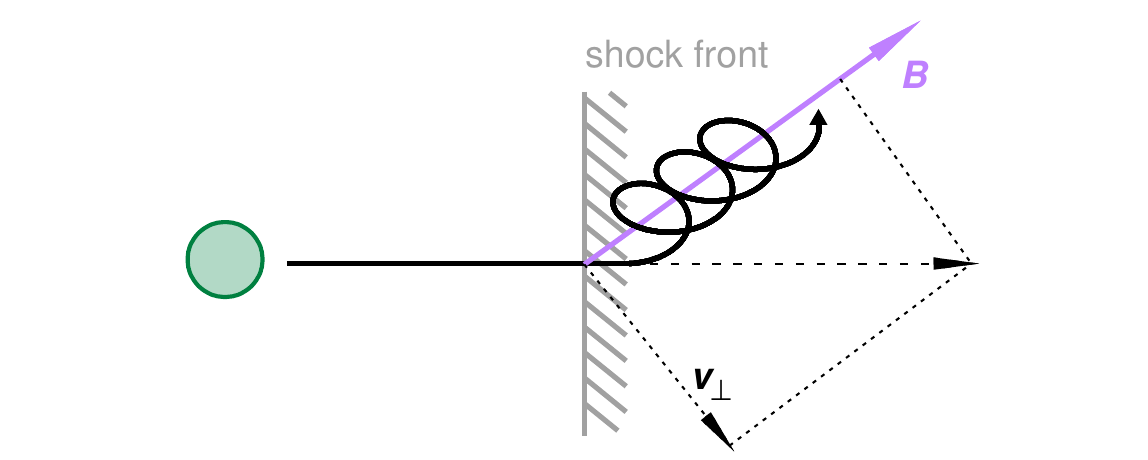}}
\caption{Sketch of a charged dust grain in the rest frame of the shock. The grain is deflected by the magnetic field from its initial propagation direction. The Lorentz force causes the gyration motion of the dust grain around the magnetic field lines.\label{sketch_Lorentz}}
\end{figure}

$\Delta \mathbf{r}$ has to be rotated back into the frame of the cloud-crushing setup by multiplying with the inverted rotation matrix $\mathbf{M}^{-1}$. Finally, the dust transport is determined for each of the sub-intervals of $\Delta t$ and the total transport of the grain during $\Delta t$ is calculated by summing up the contributions of each sub-interval (equation~\ref{eq_transport}). The transport velocity during $\Delta t$ is then 
\begin{align}
 \Delta \mathbf{v}_\text{trans} = (\mathbf{r}(t +\Delta t) - \mathbf{r}(t))/\Delta t. \label{32e1}
\end{align}
Fig.~\ref{sketch_Lorentz} shows a sketch of a charged dust grain entering orthogonally the shock-front. Without magnetic field, the grain will continue to move in the same direction as before (dashed line). The gas drag will cause a slowing down of the grain and after a while, the grain is at rest compared to the shocked gas.  However, considering a magnetic field the grain will be deflected and starts to gyrate around the magnetic field lines. The initial gyration velocity is equal to the grain velocity component
 perpendicular to the magnetic field when entering the shocked region. The grain also moves parallel to the magnetic field lines with an initial velocity that is equal to the parallel component of the grain velocity. Gas drag will continuously reduce both the dust velocity and the gyration radius. Finally, the dust grain will be at rest, fully coupled to the shocked gas.  
 \subsection{Dust gyration} 
 
      \begin{figure*}
 \resizebox{\hsize}{!}{ 
  \includegraphics[trim=2.2cm 2.2cm 0.3cm 2.05cm, clip=true,  page=1]{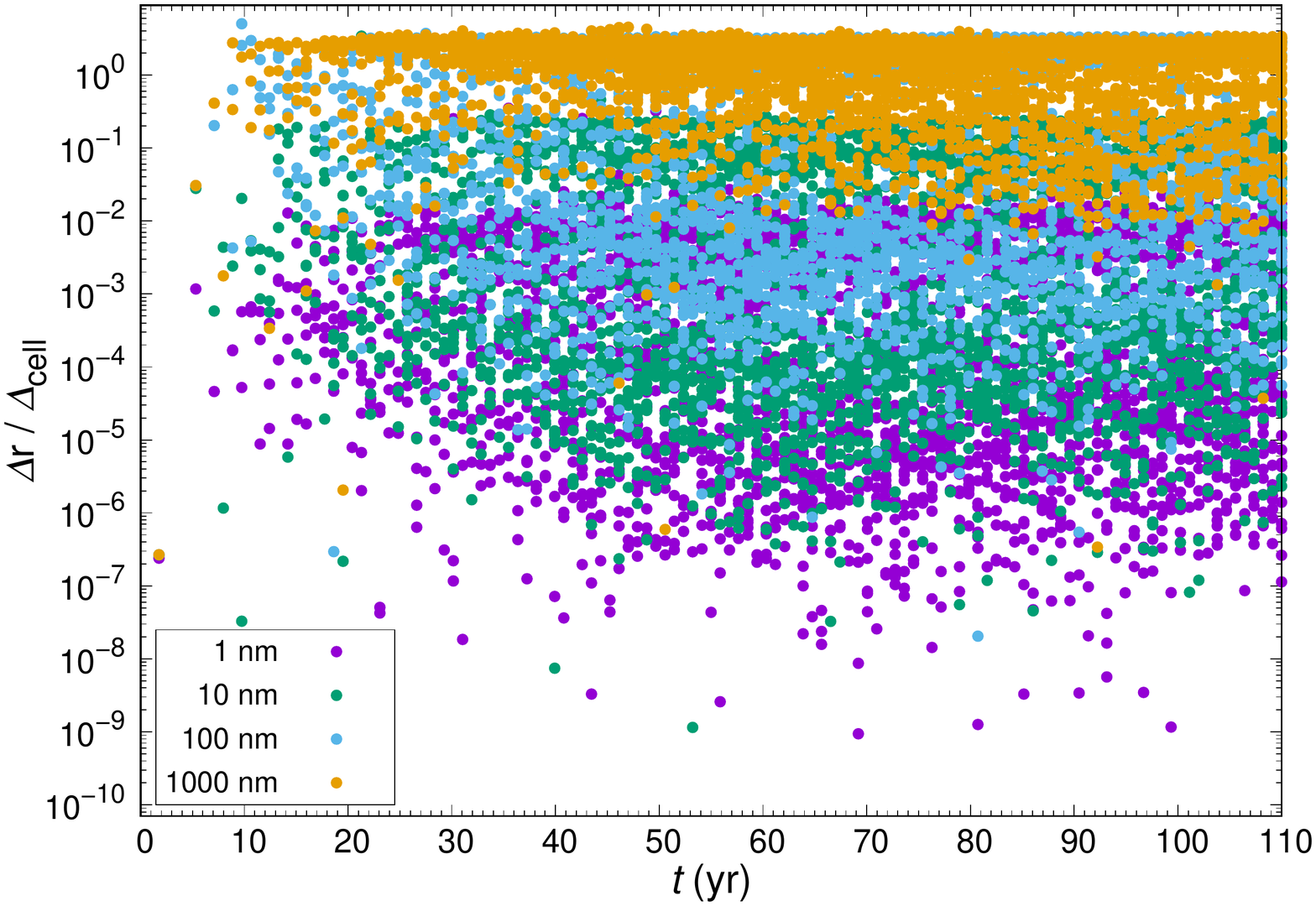}
  \includegraphics[trim=2.2cm 2.2cm 0.3cm 2.05cm, clip=true,  page=1]{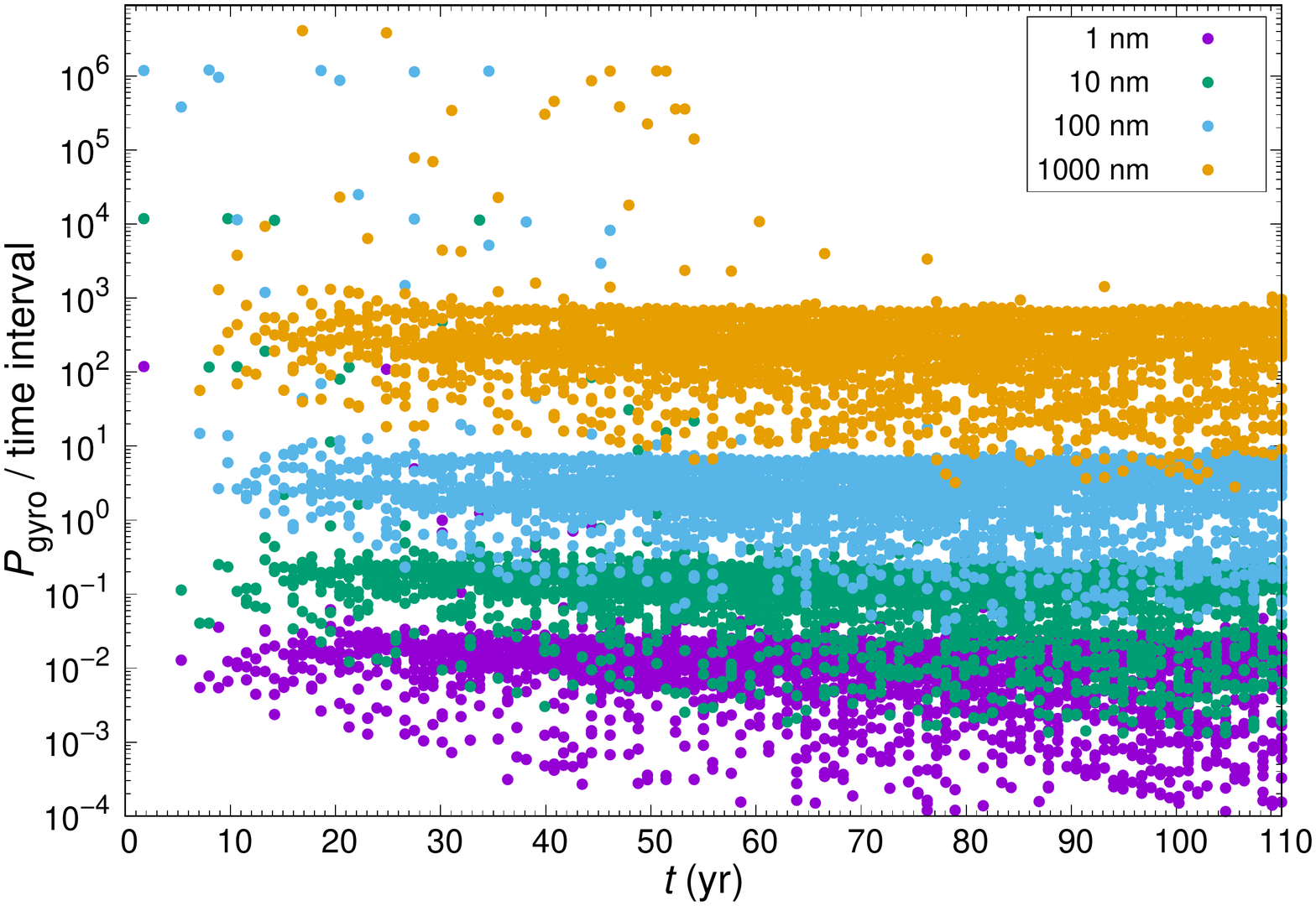}
  }
  \caption{\textit{Left}: Change of position (displacement) $\Delta r$ during the smallest time interval divided by the physical resolution of each grid cell $\Delta_\text{cell} = 10^{14}$cm for an example simulation with density contrast $\chi=300$ and magnetic field $B_0=1\,\mu$G perpendicular to the shock direction. The different colors represent the grain sizes $1\,$nm, $10\,$nm, $100\,$nm, and $1000\,$nm. Dust destruction or grain growth processes are ignored. For a better visualization, the results of only $0.1$~per~cent (randomly chosen) of all cells are shown. \textit{Right}: Gyration period divided by the smallest time interval ($0.089\,$yr).\label{fig_gyrationradius_and_period}}
  \end{figure*}
 
 When a charged dust grain moves relative to a magnetic field, the grain is deflected from a linear movement and instead gyrates around the magnetic field lines. Depending on the size of the cells in the domain and the time interval during which the movement is calculated, the charged grains are potentially able to leave the cell in which they started. Fig.~\ref{fig_gyrationradius_and_period} (\textit{left}) shows the displacement of the dust grain due to the Lorentz force during the smallest time interval ($\Delta t/10$) divided by the physical resolution of each grid cell $\Delta_\text{cell} = 10^{14}$cm. We do not show the gyration radius because the change of position also depends on the gyration period, while the displacement shows the true change of position. The displacement is shown for an example simulation with the density contrast $\chi=300$ and a magnetic field strength $B_0=1\,\mu$G perpendicular to the shock direction. The simulation is without dust destruction and without dust growth processes and represents the four dust grain radii $1\,$nm, $10\,$nm, $100\,$nm, and $1000\,$nm. We can see that in most cases the displacement is much smaller than the physical resolution of the grid cell, however, in particular for the large grains ($100\,$nm and $1000\,$nm) the change of position is in many cases comparable or even larger than the extension of a single grid cell. As a consequence, the dust grain leaves the cell and moves into another cell of the domain. In the following time step, the grain charge is calculated on the basis of the gas conditions in this new cell. The dust grain velocity is calculated on the basis of both the gas conditions in this new cell and on the previous dust velocity. In general, this procedure is the same as for the grain dynamics in an unmagnetized gas (\citetalias{Kirchschlager2019b}) where grains can leave the cell due to advection in a flowing (shocked or turbulent) gas. For details for assigning grains to spatial grid cells (and in the case of dust destruction or grain growth also assigning grains to the grain size bins), we refer to Section~4.7 in \citetalias{Kirchschlager2019b}.
 
Fig.~\ref{fig_gyrationradius_and_period} (\textit{right}) shows the gyration period divided by the smallest time interval of $0.089\,$yr for the same example simulation.  We see that for most of the large grains ($100\,$nm and $1000\,$nm), the gyration period is very well resolved ($P_\text{gyro}/\text{time interval}\gg1$), while the period of small grains is in most cases unresolved ($P_\text{gyro}/\text{time interval}\lesssim1$). However, this is not a problem for the calculations of the dust dynamics. Using equation~(\ref{eq_change}), the exact position of the dust grain after a time step is calculated assuming that the grain is moving on a perfect spiral instead of evolving  the spiral movement by itself following e.g. an Eulerian approach. Therefore, a temporally unresolved gyration movement is no issue to determining the position of a dust grain after a specific amount of time.


 \subsection{Grain charging} 
The dust grains need to be charged to be affected by the Lorentz force. Several quantities and processes influence the total charge of the grain such as the kind of impinging plasma particles, associated secondary electrons, transmitted plasma particles, and field emission. As in \citetalias{Kirchschlager2019b} we apply for our dust-processing simulations the analytical description of the charging processes summarized in \cite{Fry2018, Fry2020}, which is based on approaches introduced by \cite{Shull1978} and \cite{McKee1987}. The grain charge $Q_\text{grain}$ depends on the gas temperature $T_\text{gas}$, the grain size $a$, the relative velocity between dust and gas as well as the gas species, and is calculated for each grain species, time-step and cell in the domain. Note, that photoelectric emission is ignored. For further details, we refer to Appendix~A in \citetalias{Kirchschlager2019b}.

\subsection{Dust destruction}
\textsc{Paperboats} simulates the dust destruction processes thermal and kinetic (non-thermal) sputtering (e.g.~\citealt{Barlow1978,Tielens1994}) as well as fragmentation and vaporization in grain-grain collisions (e.g.~\citealt{Jones1994, Hirashita2009}). Kinetic sputtering, fragmentation and vaporization are processes that depend on the actual velocity of the dust grains at a specific time $t$ and not on an average velocity during a time interval $\Delta t$. Therefore, the actual dust velocity $\mathbf{v}(t)$ is needed which is given by
\begin{align}
\mathbf{v}_\text{act}(t) =
\left(
  \begin{array}{l}
v_\text{perp} \cos{(\omega\,t)}\\
v_\text{perp} \sin{(\omega\,t)}\\
v_\text{para}
  \end{array}
  \right). \label{eq_actveloc}
\end{align}
For kinetic sputtering, equation~(\ref{eq_actveloc}) is directly used to calculate the sputtering yields of a dust grain in a moving gas. For fragmentation and vaporization, the difference between the actual velocity $\mathbf{v}_\text{act}$ of grains of different sizes is calculated and applied to the grain-grain collision routine as outlined in \citetalias{Kirchschlager2019b}. In general, the relative velocities between the dust grains and the gas as well as between grains of different sizes are increased as they are not anymore moving in the same direction.  The higher relative velocities have then the potential to cause i) a higher number of grain-grain collisions (or for sputtering of a dust grain: more collisions with gas particles), and ii) a higher destruction rate at a single collision as the collision velocity is higher.

\subsection{Partial grain vaporization}
\label{400d}

In the original version of \textsc{\mbox{Paperboats}} (\mbox{\citetalias{Kirchschlager2019b}}), vaporization of dust grains (total destruction of a grain in a grain-grain collision) takes place when the collision velocity of two dust grains is above a certain velocity threshold. For silicates, e.g., this threshold is at $v_\text{vapo}=\unit[19]{km/s}$. The collision of two dust grains of arbitrary size at a higher velocity causes then the total destruction of both grains. This implies an unrealistic behaviour if one of the grains is much bigger than the other, e.g., a $\unit[1]{nm}$ particle impacting a 1 micrometre grain with a velocity $v> \unit[19]{km/s}$ will automatically lead to vaporization of both grains, and there is no chance that only the small grain is vaporized while the big one can at least partially survive.

In order to treat grain vaporization in a more realistic way, we follow the idea of \cite{Borkowski1995}, use the binding energy of the grain atoms and compare it to the collision energy. 
The effective binding energy of a grain atom is $E_\text{vap} = 0.74\left\langle M_\text{atom}\right\rangle\text{eV}$, where $\left\langle M_\text{atom}\right\rangle$ is the average atomic mass of the grain atoms in atomic mass units $m_\text{amu}$. For silicate $\left\langle M_\text{atom}\right\rangle=20$ and for carbon $\left\langle M_\text{atom}\right\rangle=12$ (\citealt{Tielens1994, Nozawa2006}). The collision energy of two grains with masses $m_1$ and $m_2$ colliding with relative velocity $v_\text{col}$ is $E_\text{col} = \frac{1}{2}\frac{m_1 m_2}{m_1+m_2} v_\text{col}^2$. Below the threshold velocity $v_\text{vapo}$, no dust grain is vaporized, only fragmentation can happen. Above the threshold velocity, the collision energy $E_\text{col}$ is used to partially or fully vaporize one or both dust grains. We assume, that the collision energy is split up to equal parts on the two grains. The number of atoms vaporized in each dust grain by a collision is then 
\begin{align}
N_\text{vap} = E_\text{col}/(2\, E_\text{vap}).
\end{align}
If $N_\text{vap}$ exceeds the total number of grain atoms in an individual dust grain, the dust grain is totally vaporized and the energy excess not required for the destruction is allocated to the second grain. If $N_\text{vap}$ is smaller than the total number of atoms in a grain, $N_\text{vap}$ atoms are removed from the grain and the new grain mass of the partially vaporized dust grain  is calculated by 
\begin{align}
m_{\text{new},i} = m_i -  N_\text{vap}M_\text{atom}m_\text{amu},
\end{align}
where $i$ is 1 or 2. This treatment of grain vaporization allows to simulate either two partially vaporized grains, or to fully vaporize the small grain while the big grain is only partially vaporized. For the present study, we adopt this grain vaporization approach.

\subsection{Dust growth}
Besides the dust destruction processes, three grain growth processes were also considered: The coagulation
of dust grains in a grain-grain collision (sticking), ion trapping\footnote{Gas particles penetrate into the dust grains and
can be trapped if the ion impact energy is sufficiently high (\citealt{Kirchschlager2020}).} of regular gas and dusty gas\footnote{The material of the dusty gas is completely atomic and composed of destroyed dust grain material (\citetalias{Kirchschlager2019b}). It is not subject to grain-grain collisions or sputtering, but contributes to grain growth processes like ion trapping and gas accretion by the surviving dust grains.}, and the accretion of regular gas and dusty gas onto the surfaces of the grains. 

Sticking has a negligible effect on the dust processing, as the present grain velocities are mostly too high (\citetalias{Kirchschlager2019b}). This is rather aggravated under the influence of a magnetic field as the velocities tend to be even larger. For the ion trapping and the gas accretion, we considered only the accumulation of the dusty gas, but not the accumulation of the regular (oxygen) gas which is subject to the MHD simulations. The accumulation of regular gas would increase the total amount of dust present in the simulation and reduce the gas mass. Due to the nature of the post-processing we are not able to reduce the gas number density from one time-step to the next which would make it difficult to regulate the formation of new grain mass (\citealt{Kirchschlager2020}). In our current simulations the accumulation of dusty gas atoms or ions back onto the dust from which they have been stripped is not found to be a significant process. In future work we plan to take into account grain growth processes caused by the accretion and ion-trapping of atoms from the gas-phase on the fly in the \textsc{\mbox{AstroBEAR}} code. 

Impinging gas particles are able to sputter grain atoms if their energy is above the threshold energy $E_\text{sp}$ (\citealt{Bohdansky1980, Tielens1994}). For lower energies, we assume the gas particles are accreted onto the dust grains with a probability $P$. In the original version of \textsc{\mbox{Paperboats}}  $P$ is proportional to the energy $E$ of the gas particle, $P(E)\propto (1 - E/E_\text{sp})$ (see equation~39 in \citetalias{Kirchschlager2019b}). We reviewed these gas sticking probabilities following \cite{Burke1983} who found 
 \begin{align}
  P(E) \sim \begin{cases}
       1                  &\text{if}\, E\le E_\text{ads},\\
    (E_\text{ads}/E)^2&\text{if}\, E> E_\text{ads}. 
   \end{cases}\label{con_Eabs}
 \end{align}
Equation~(\ref{con_Eabs}) is independent of the sputtering threshold energy $E_\text{sp}$ which is more realistic. The adsorption energy $E_\text{ads}=1.45\,$eV is taken from \cite{Molpeceres2019} and represents oxygen particles colliding with fosterite grains, but one would expect similar behavior for other heavy element atoms, e.g.,~Mg or Si~atoms. 

We note that neither of the dust growth processes have a crucial impact on the results in Section~\ref{450}. 
 

%

\subsection{Dust properties}
Some theoretical studies of dust formation predict grain size distributions in SN ejecta that can be approximated by a lognormal function (e.g.,~\citealt{Nozawa2003}). 
We adopt this approximation and assume lognormal grain size distributions in the unshocked clump. We vary the grain radius $a_\text{peak}$, at which the lognormal size distribution has its maximum, between $\unit[1]{nm}$ and  $\unit[5]{\mu m}$ to take into account grain size predictions from both observations ($\unit[{\sim}0.1]{\mu m}$ up to a few micrometers; \citealt{Gall2014, Fox2015, Wesson2015, Bevan2017, Priestley2019b, NiculescuDuvaz2022}) as well as from nucleation and coagulation theory ($\unit[{\sim}1]{nm}$ up to a few $\unit[100]{nm}$; \citealt{Todini2001, Nozawa2003, Bocchio2014, Sarangi2015, Biscaro2016}; \citealt{Sluder2018}). In total, we consider 12 different values for $a_\text{peak}$. The initial grain size distribution width is fixed to $\sigma=0.1$ and represents a relatively narrow size distribution. The grains of the size distribution are binned in 40 log-spaced bins ranging from $\unit[0.6]{nm}$ up to $\unit[10]{\mu m}$ for the grain size distribution with $a_\text{peak} = \unit[5]{\mu m}$, and in 32 log-spaced bins ranging from $\unit[0.6]{nm}$ up to $\unit[3]{\mu m}$ for the 11 grain size distributions with  $a_\text{peak} < \unit[5]{\mu m}$. Grains processed by sputtering or grain-grain collisions with radii below $\unit[0.6]{nm}$ are considered as obliterated dust masses (dusty gas) and no longer as dust grains. Additionally to the 32 or 40 grain size bins, we follow the dusty gas in a collector bin ($a<\unit[0.6]{nm}$) and monitor dust grains that have grown to sizes above the maximum grain radius (3 or $\unit[10]{\mu m}$) in a spatially unresolved upper bin. The dust grains are composed of either silicate or carbon. The full set of material properties required for the dust post-processing is given in Table~2 of \mbox{\citetalias{Kirchschlager2019b}}.

As the dust grains are assumed to have formed in the over-dense gas clumps in the ejecta (\citealt{Lagage1996, Rho2008, Lee2015}), dust is present in the pre-shock clump while the pre-shock ambient medium is dust-free. Within the spherical clump, the dust is homogeneously distributed with a gas-to-dust mass ratio of 10 (\citealt{Priestley2019a}).  Based on the gas number density, gas-to-dust mass ratio and the dust grain size distribution at the beginning of the simulation, the number density of each of the 32 (40) dust grain sizes is calculated for each cell in the domain. Subsequently, the number density of each bin and cell is calculated for later time-points by considering changes due to dust transport and grain destruction or growth.

We do not consider any feedback of the dust to the gas because of the nature of the post-processing. We neglect the addition of coolants to the gas from the destroyed dust material as well as the momentum transmission from the grains to the gas by the gas or plasma drag. 

\section{Results}
\label{450} 

\subsection{MHD evolution of the shocked gas}
\label{s510} 

    \begin{figure}
 \resizebox{\hsize}{!}{
   \includegraphics[trim=8.0cm 1.5cm 1.9cm 0.0cm, clip=true, page=1]{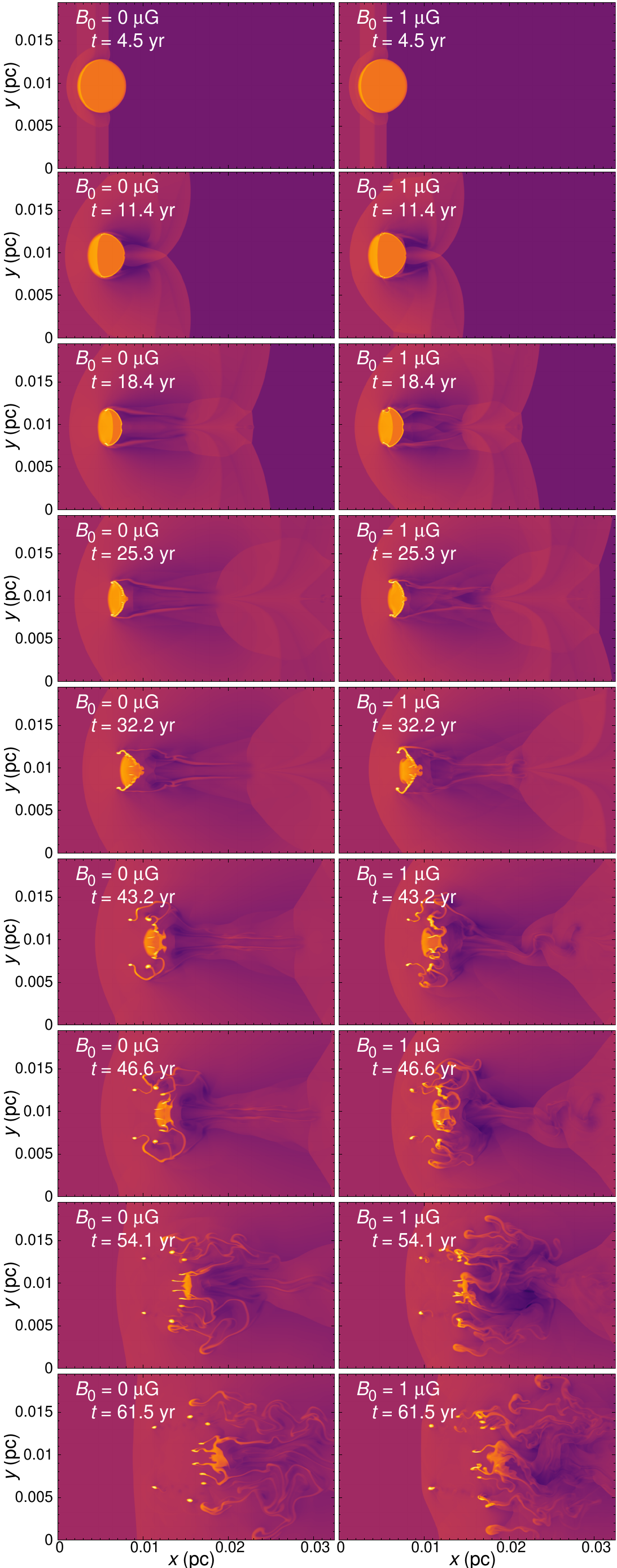}
   }\\[-0.5cm]
 \caption{Temporal evolution (from \textit{top} to \textit{bottom}) of the spatial gas density without (\textit{left}) or with (\textit{right}) initial magnetic field ($B_0=1\,\mu$G). The density contrast is $\chi = 100$. At $t=0$ the shock front enters the computational domain and hits the clump after ${\sim}2\,$yr. The panels show a fixed cut-out of the computational domain and the colour scale is fixed. Click \href{https://youtu.be/ufjy58X2xHo}{here} to play a short movie showing the temporal evolution of the magnetic field case. The movie frame rate is set to $10$ frames s${}^{-1}$.\label{gas_chi=100}}
 \end{figure}
   
 \begin{figure}
 \resizebox{\hsize}{!}{
   \includegraphics[trim=8.0cm 1.5cm 1.9cm 0.0cm, clip=true, page=1]{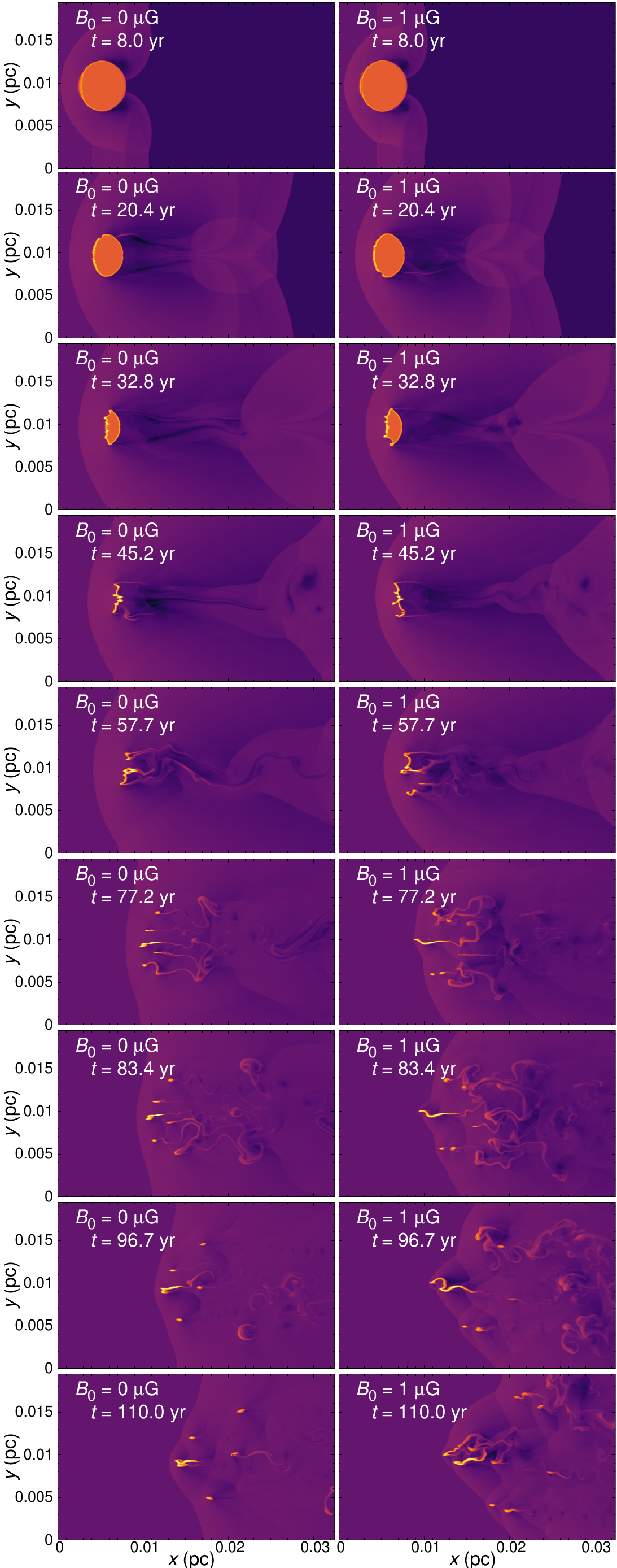}
   }\\[-0.5cm]
 \caption{Same as Fig.~\ref{gas_chi=100}, but for a density contrast $\chi = 300$. Click \href{https://youtu.be/6ZmppLkGHFw}{here} to play a short movie showing the temporal evolution of the magnetic field case.\label{gas_chi=300}}
 \end{figure} 
 
   \begin{figure*}\vspace*{-0.3cm}
  \resizebox{0.915\hsize}{!}{
    \includegraphics[trim=1.1cm 5.4cm 1.0cm 8.1cm, clip=true, page=2]{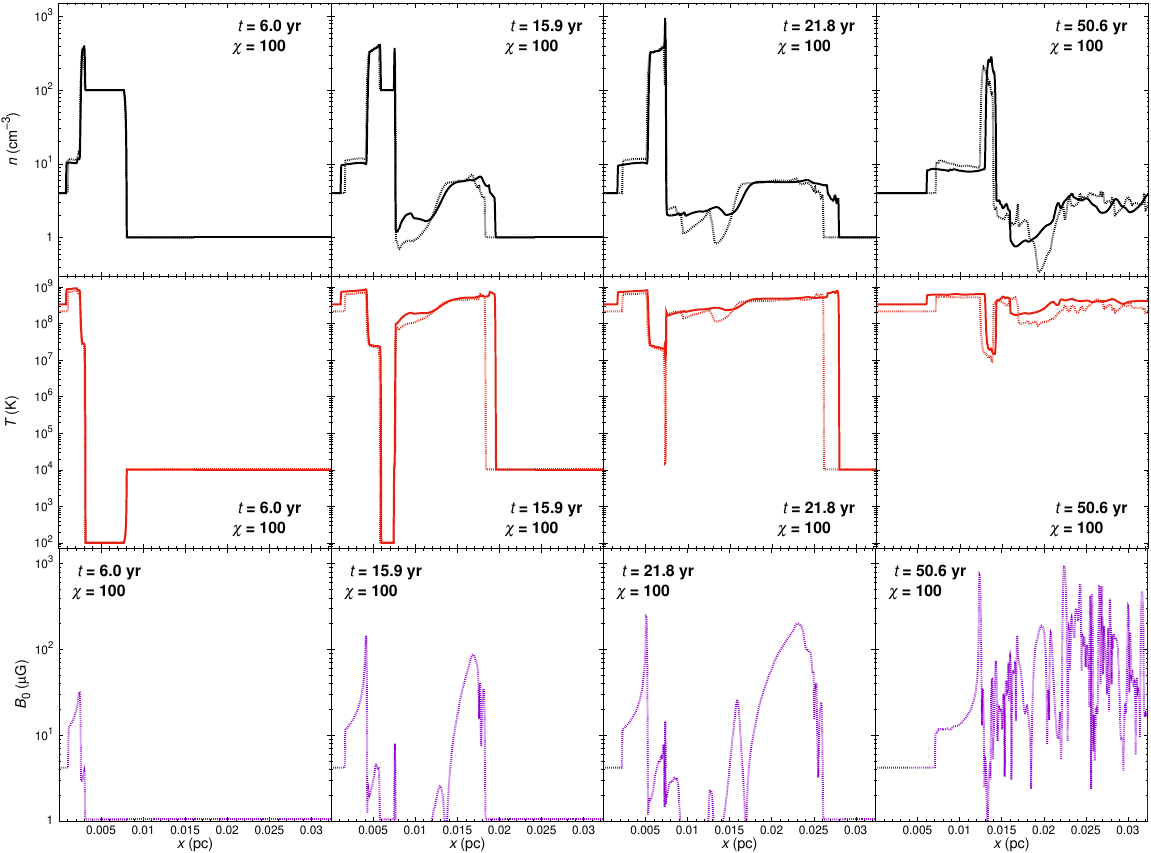}
    }
  \resizebox{0.915\hsize}{!}{    
    \includegraphics[trim=1.1cm 5.4cm 1.0cm 8.1cm, clip=true,page=3]{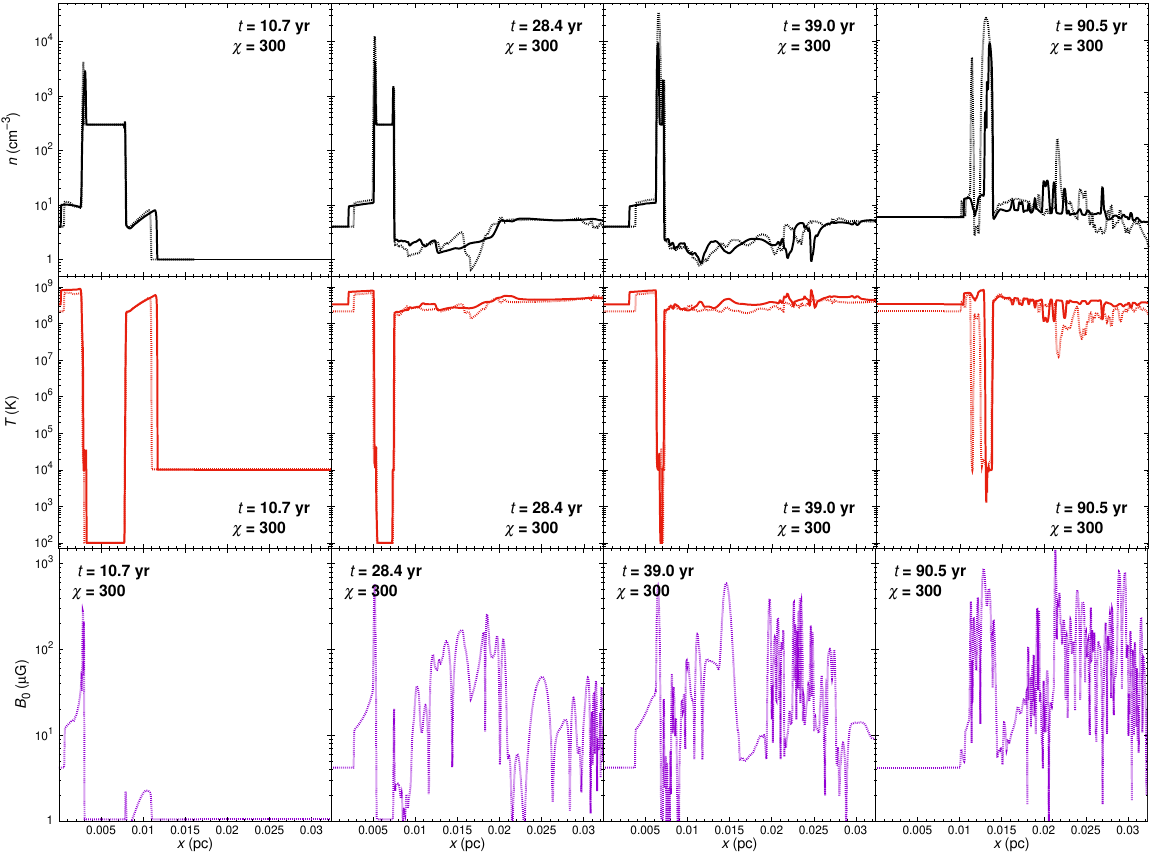} 
    }\\[-0.2cm]
  \caption{Profiles of the gas density, gas temperature, and magnetic field strength for a density contrast $\chi = 100$ (\textit{top}) and $\chi = 300$ (\textit{bottom}). The $x$-axis represents the shock direction through the mid-point of the original clump. Simulations with $B_0=0$ are shown as solid lines, dotted lines represent $B_0=1\,\mu$G.\label{cuts_chi=100or300}}
  \end{figure*}

     \begin{figure*}
 \resizebox{\hsize}{!}{
   \includegraphics[trim=0cm 12.7cm 0cm 0cm, clip=true, page=1]{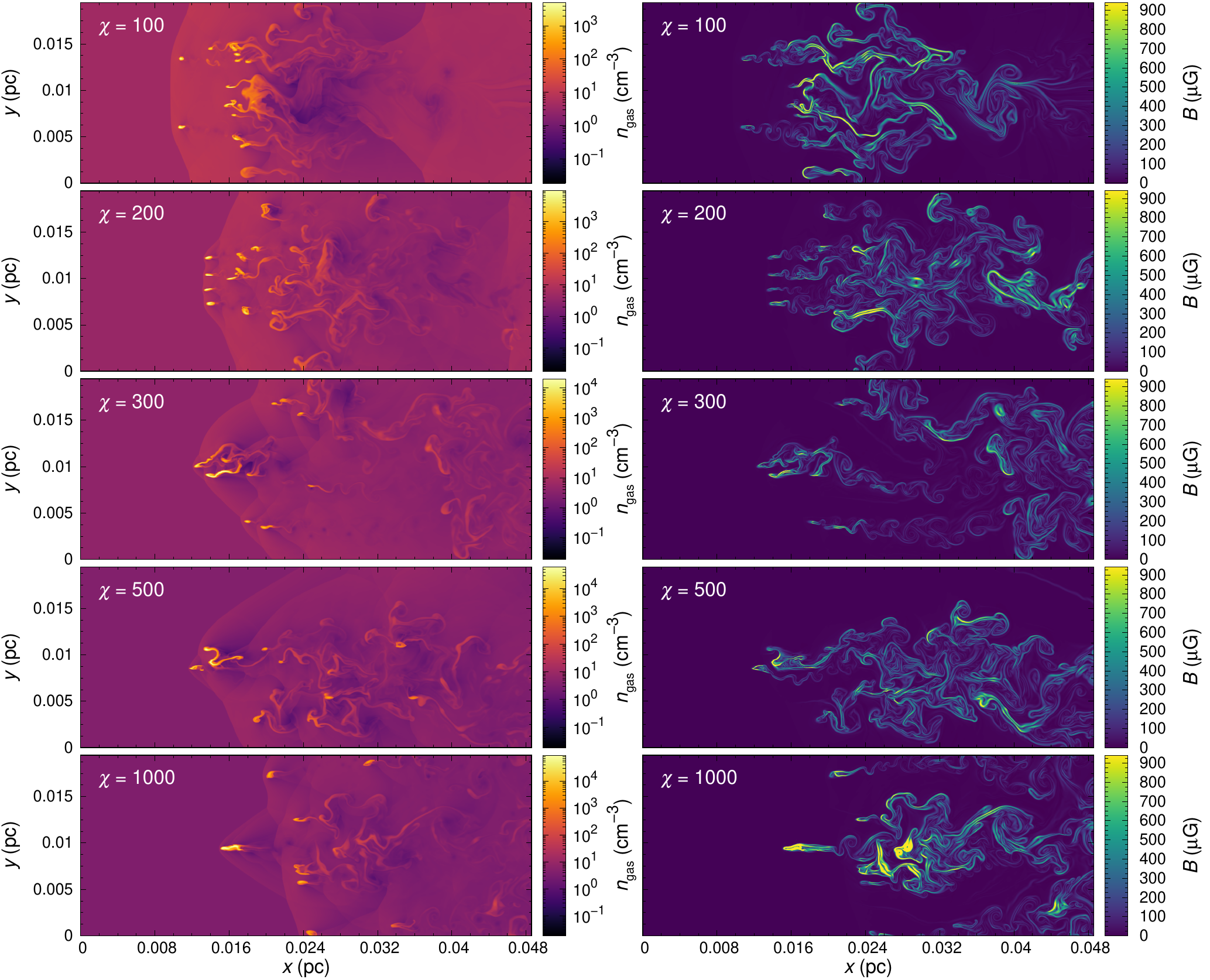}
   }
 \caption{Impact of the density contrast (from top to bottom: $\chi=100$, 200, 300, 500, 1000) on the gas density (\textit{left}) and magnetic field strength (\textit{right}) after three cloud-crushing times. The initial magnetic field strength is for all cases $B_0=\unit[1]{\mu G}$.\label{gas_chi}}
 \end{figure*}
 
    \begin{figure*}
 \resizebox{\hsize}{!}{
   \includegraphics[trim=0cm 12.8cm 0cm 0cm, clip=true, page=1]{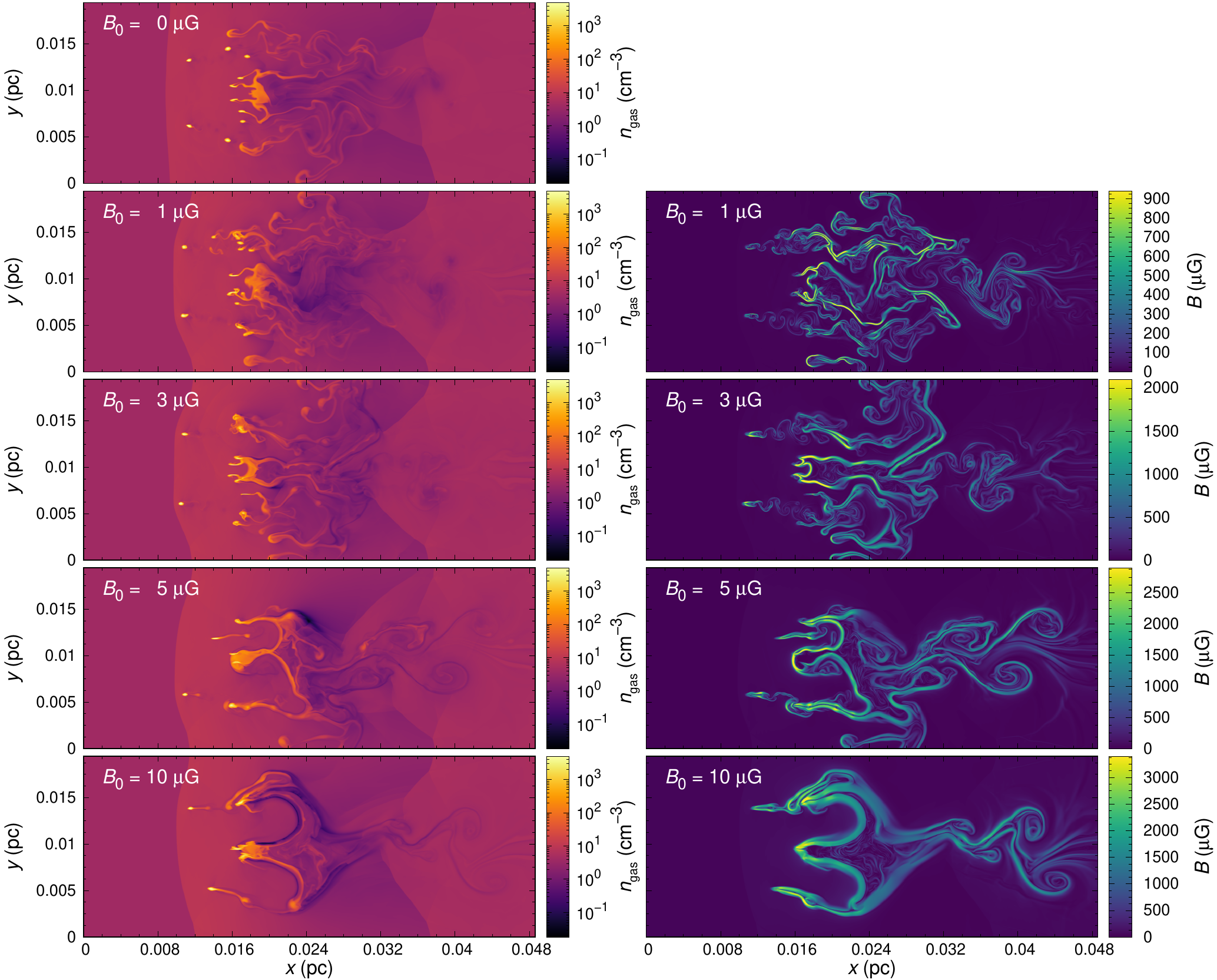}
   }
 \caption{Impact of the initial magnetic field strength (from top to bottom: $B_0=0$, 1, 3, 5, $\unit[10]{\mu G}$) on the gas density (\textit{left}) and magnetic field strength (\textit{right}) after three cloud-crushing times. The density contrast is for all cases $\chi=100$.\label{gas_magn}}
 \end{figure*}

   \begin{figure*}
 \resizebox{\hsize}{!}{
   \includegraphics[trim=0cm 12.8cm 0cm 0cm, clip=true, page=1]{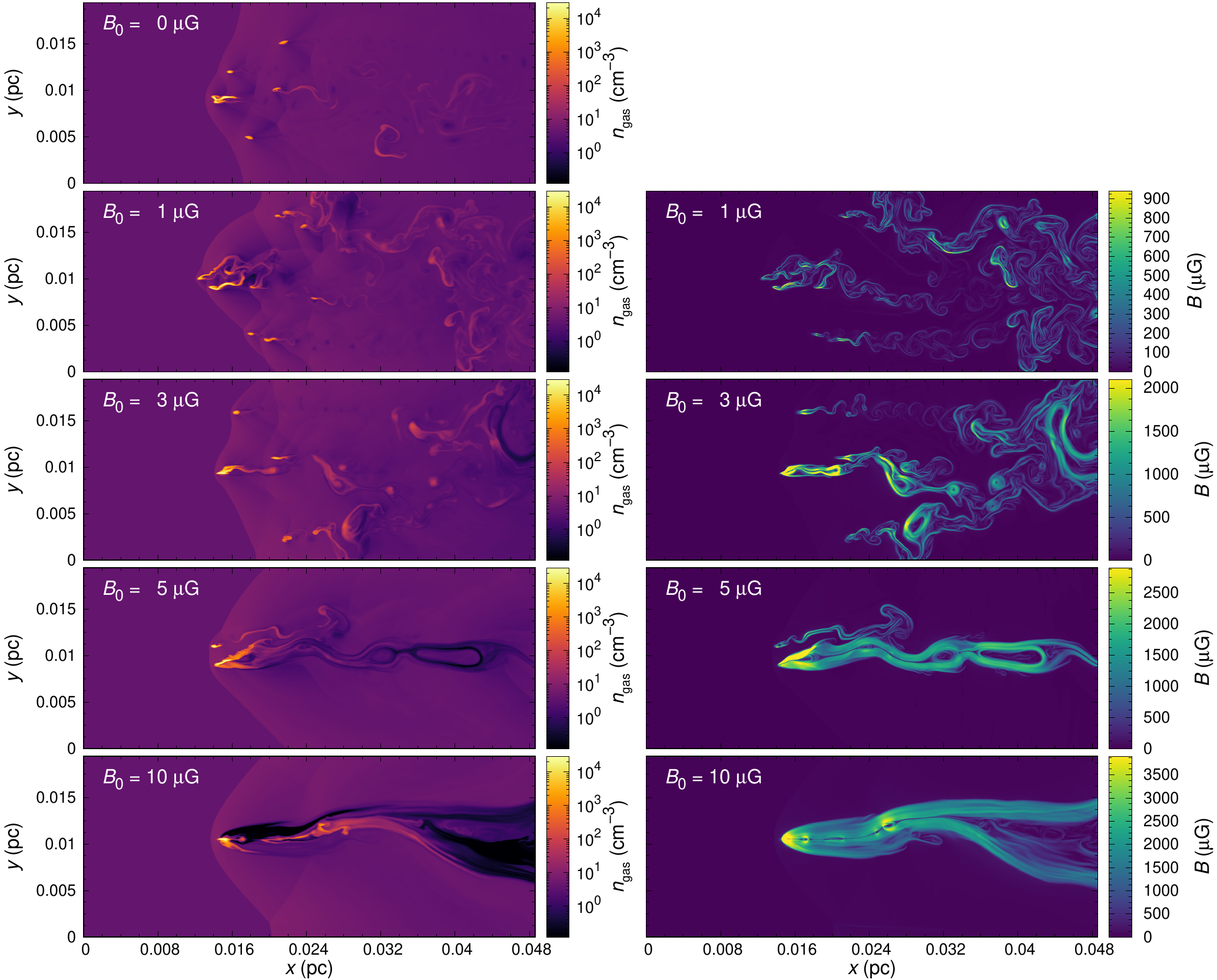}
   }
 \caption{Same as Fig.~\ref{gas_magn}, but for a density contrast $\chi = 300$.\label{gas_magn_chi=300}}
 \end{figure*}

The disruption of a clump of gas embedded in the ambient medium of the ejecta is shown in Figs~{\ref{gas_chi=100}-\ref{gas_magn_chi=300}} for different density contrasts. 

In Figs~\ref{gas_chi=100} and \ref{gas_chi=300} we follow the temporal evolution of the gas number density when the initial magnetic field strength is $B_0=0\,\mu\text{G}$ or $1\,\mu\text{G}$ and the density contrast $\chi=100$ or 300. The reverse shock impacts the clump, compresses it, strips off material from the outer shells of the clump and blows it away, and finally accelerates and fragments the clump, forming high-fractal, non-symmetric structures. Compared to \citetalias{Kirchschlager2019b}, the maps show significantly finer and smaller structures which is solely due to the finer grid in this study (resolution increased by a factor of 5). Though the gas density maps with or without magnetic field are not identical, no qualitative differences can be determined. Note, that the time required to disrupt the clump is longer for the density contrast $\chi=300$ compared to $\chi=100$ due to the longer cloud-crushing time ($\tau_\text{cc}=34.3\,$yrs vs. $19.81\,$yrs). 

The corresponding radial profiles of the gas density, gas temperature, and magnetic field strength are presented in Fig.~\ref{cuts_chi=100or300}. The snapshots are at ${\sim}0.2\,\tau_\text{cc}$, $0.7\,\tau_\text{cc}$, $1.0\,\tau_\text{cc}$, and $2.45\,\tau_\text{cc}$ after the first contact
of the shock with the clump and can be compared to fig.~5 in \cite{Silvia2010} and to fig.~18 in \citetalias{Kirchschlager2019b}. Compared to the latter, the profiles of the new simulations show also finer structures but agree otherwise. Differences between the simulations with or without magnetic field are visible but not crucial.

There are two parameters which significantly influence the evolution of the shocked gas: the density contrast $\chi$ and the strength of the initial magnetic field $B_0$. Fig.~\ref{gas_chi} shows the gas number density and the spatial distribution of the magnetic field strength in the post-shock gas for different density contrasts $\chi$. The initial magnetic field strength is fixed to $B_0=1\,\mu\text{G}$ for all cases. The higher the initial density contrast, the higher are the densities in the knots of the shocked gas. Though the final gas distribution depends on the density contrast $\chi$, no clear trend for the occurrence of structures and instabilities is visible. For the density contrasts $\chi=300$ and 500 the extent of the shocked clump material is squeezed in the vertical direction (perpendicular to the shock direction) compared to smaller ($\chi=100$ and 200) or higher density contrasts ($\chi=1000$). The magnetic field strength (Fig.~\ref{gas_chi}, \textit{right}) is amplified by the shock compression from an initial value of $B_0=1\,\mu\text{G}$ to a few $100\,\mu\text{G}$. The larger the density contrast, the larger is the maximum magnetic field strength that can be detected in the shocked gas. The same structures and instabilities are visible in the maps of the gas density and in the magnetic field maps though the regions with high gas densities do not coincide inevitably with regions of high magnetic fields strengths. The magnetic fields are easily dragged by the gas motion, the flows compress the gas and the magnetic field lines are reinforced leading to the higher magnetic field strengths. However, the magnetic field lines are not perfectly frozen to the gas.
 
Figs~\ref{gas_magn} and \ref{gas_magn_chi=300} show the gas number density and the spatial distribution of the magnetic field strength after three cloud-crushing times for different initial magnetic field strengths $B_0$.  The density contrast is again $\chi=100$ or 300. The initial magnetic field has a crucial impact on the final gas distribution. The smaller $B_0$, the finer are the structures in the maps. For $B_0\gtrsim 5\,\mu\text{G}$, the shocked clump consists only of a few stretched fragments which are partly connected. There is no indication that a higher initial magnetic field causes denser knots in the shocked gas. The magnetic field strength (Figs.~\ref{gas_magn} and \ref{gas_magn_chi=300}, \textit{right}) of the shocked gas, on the other hand, sensitively depends on the initial magnetic field strength $B_0$. The larger the initial magnetic field, the larger are the maximum magnetic field strengths in the shocked gas. The magnetic field maps show again similar structures and instabilities as the gas density maps without the agreement of the maxima or minima positions of the gas densities and the magnetic field.

In conclusion, the simulations of all density contrasts and initial magnetic field strengths show that the clump is mostly disrupted after three cloud-crushing times and the final fragments are distributed as diffuse material in the shocked gas. 

\subsection{Coupling between gas and dust}
\label{511a}  
\begin{figure*}
 \resizebox{\hsize}{!}{
   \includegraphics[trim=2.8cm 6.3cm 1.1cm 5.4cm, clip=true, page=1]{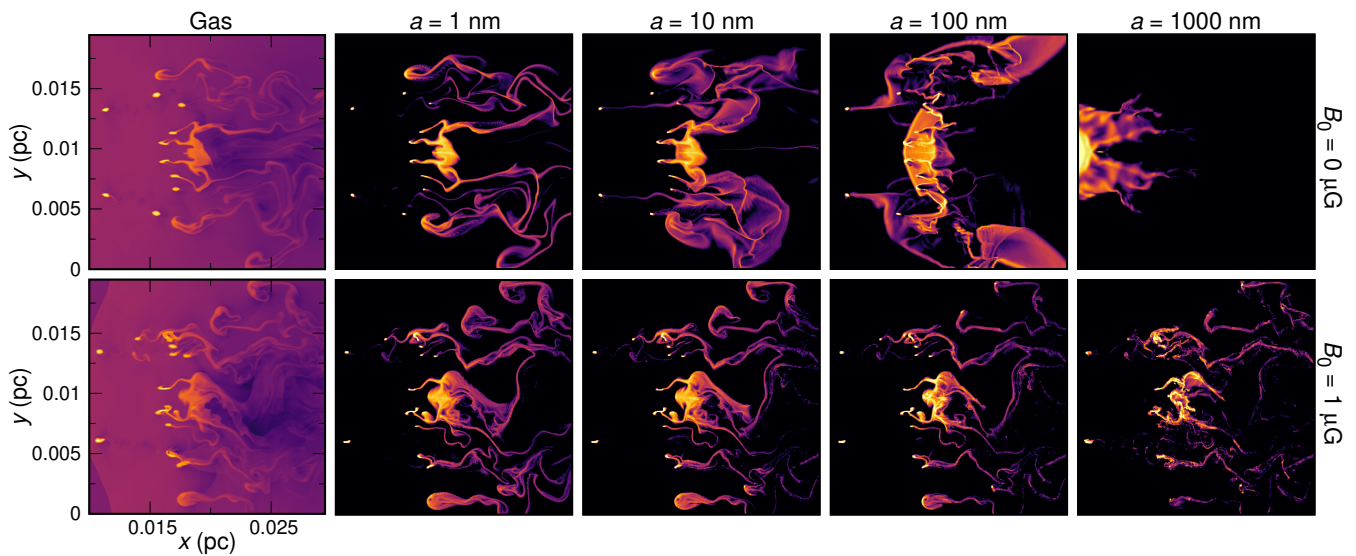}
   }
 \caption{The density distributions of the fragmented clump (columns from \textit{left} to \textit{right}: gas number density, dust number density for the single grain sizes $a=1\,$nm,  $10\,$nm, $100\,$nm, and $1000\,$nm) at ${\sim}60\,$yrs after the clump was hit by the shock wave. Each panel shows the same box of the domain and the logarithmic colour scale is fixed for each column. The density contrast of the clump is $\chi=100$. The Figure shows that dust grains are better coupled to the gas in the case of a magnetic field of $B_0=1\,\mu$G (\textit{bottom} row) compared to the non-magnetic case (\textit{top}). For both cases: The larger the grain size, the larger the decoupling of gas and dust, however, decoupling only becomes significant for grains with radii $a\gtrsim100\,$nm (for $B_0=0\,\mu$G) or $a\gtrsim1000\,$nm (for $B_0=1\,\mu$G).
 \label{fig_coupling_100}}
 \end{figure*}

\begin{figure*}
 \resizebox{\hsize}{!}{
   \includegraphics[trim=2.8cm 6.3cm 1.1cm 5.4cm, clip=true, page=1]{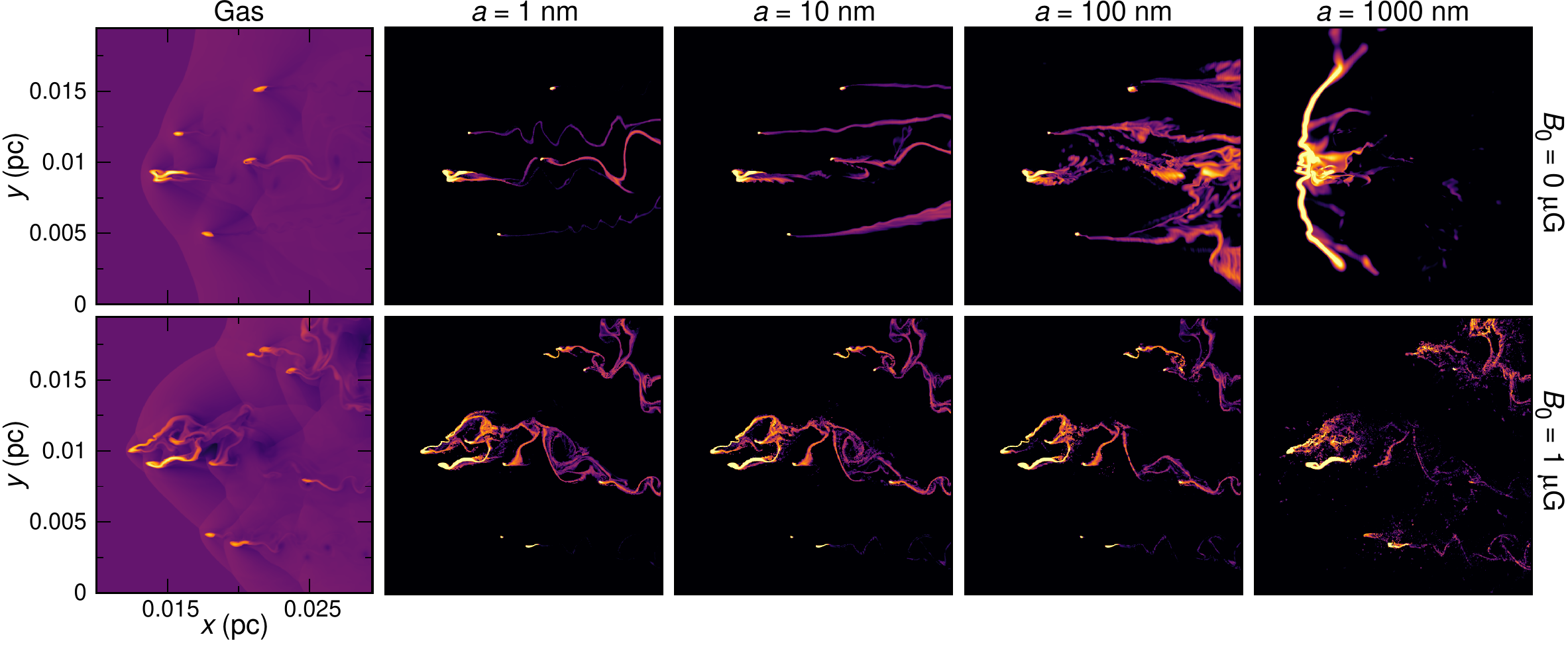}
   }
 \caption{Same as Fig.~\ref{fig_coupling_100}, but for a density contrast $\chi=300$.\label{fig_coupling_300}}
 \end{figure*}

We use our post-processing code \textsc{\mbox{Paperboats}} to calculate maps of the dust grain distributions on the basis of the MHD output of \textsc{\mbox{AstroBEAR}}. In order to study the coupling between gas and dust, we ignore in this section dust destruction and grain growth and follow the dust transport of grains with radii $a=1\,$nm,  $10\,$nm, $100\,$nm, and $1000\,$nm for a magnetic ($B_0=1\,\mu$G) and a non-magnetic case ($B_0=0\,\mu$G). We choose a specific square box in the maps which contains the main parts of the fragmented gas clump after three cloud-crushing times. Figs.~\ref{fig_coupling_100} and~\ref{fig_coupling_300} show the gas and dust density in that box for the density contrasts $\chi=100$ and 300, respectively. The differences between the maps of gas have been discussed already in Section~\ref{s510} and we focus here on the coupling or decoupling.

In all cases the $1\,$nm and the $10\,$nm grains couple very well to the gas. For larger grains, we can clearly see a difference.  In the non-magnetic case, the distributions of the $100\,$nm grains and of the gas differ which indicates a decoupling. Larger grains ($1000\,$nm) have a large inert mass and lag distinctly behind the gas flow. For the density contrast $\chi=100$ (Fig.~\ref{fig_coupling_100}, \textit{top} \textit{right}), the main part of the $1000\,$nm grains has not fully arrived at the square box after three cloud-crushing times and just appears at the edge of the box. In the magnetic case ($B_0=1\,\mu$G), the coupling between the $100\,$nm grains and the gas is clearly better. The first weak differences appear for the $1000\,$nm grains which start to decouple from the gas, however, the coupling is still strong.

In summary, the larger the grain size, the weaker is the coupling of gas and dust. The presence of magnetic fields reinforces the coupling. The differences seen between the dust maps for the density contrasts  $\chi=100$ and 300 are dictated by the differences in the maps of the gas. We note that despite a stronger coupling of gas and dust, magnetic fields can potentially increase the relative velocities  between dust grains and the gas as well as for dust grains of different sizes as the charged grains gyrate around the magnetic field lines.

 
\subsection{Dust destruction in the shocked clump}
\label{511} 
     \begin{figure*}
 \resizebox{\hsize}{!}{ 
  \includegraphics[trim=2.8cm 2.4cm 0.9cm 2.0cm, clip=true,  page=1]{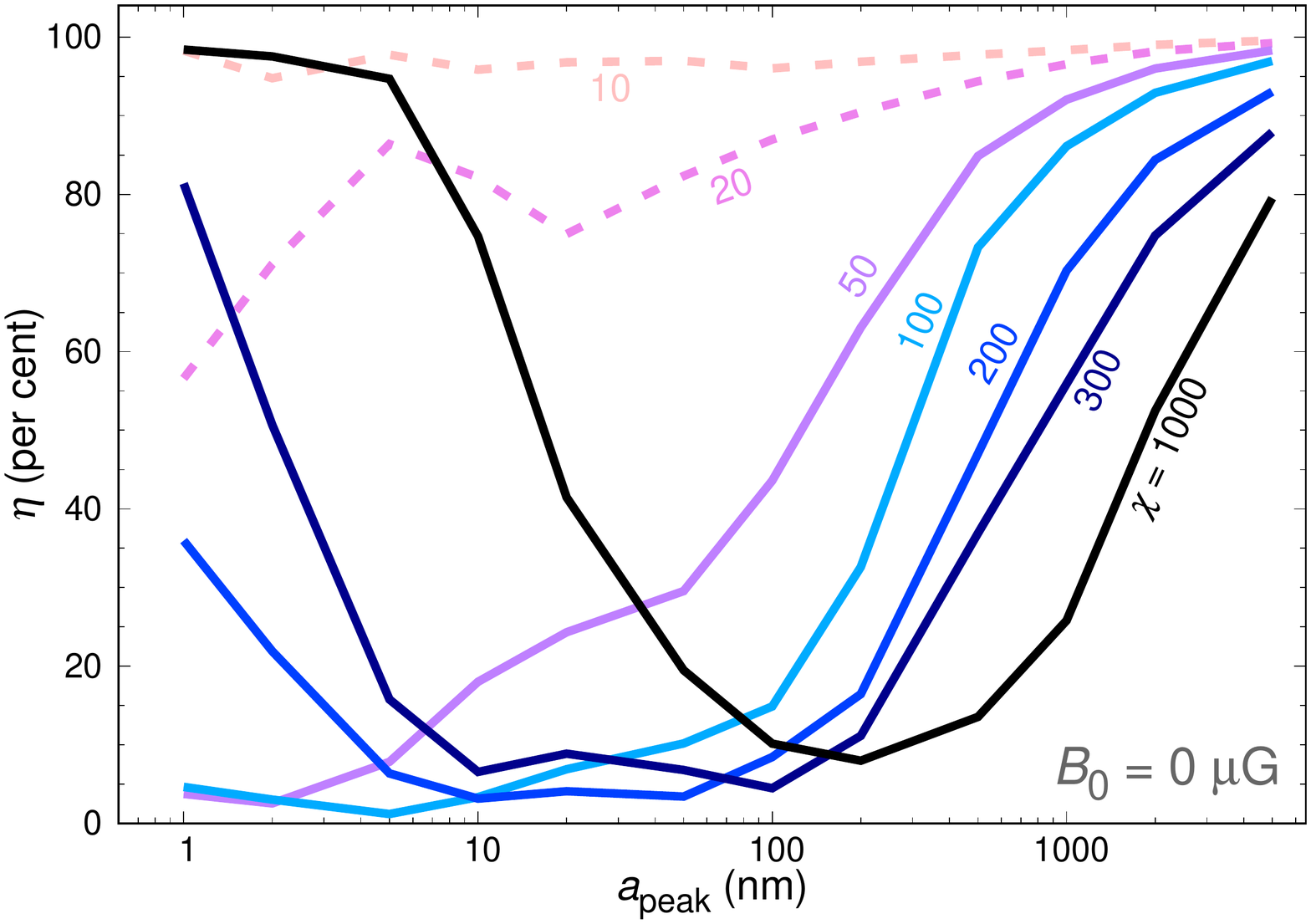}
  \includegraphics[trim=1.5cm 2.4cm 0.9cm 2.0cm, clip=true,  page=4]{Pics/Results_paper/Survival_chi.pdf}
  }
  \caption{Silicate survival fraction $\eta$ (\textit{left}) and total surviving dust mass (\textit{right}) in a clump with gas density contrast $\chi$ (coloured lines) between the clump and the ambient medium. The initial magnetic field strength is $B_0=\unit[0]{\mu G}$. The gas density contrast $\chi$ varies between 10 to 1000, however, we constrain our analysis on $\chi\ge50$ as these density contrasts are more realistic for clumps in Cas~A.\label{fig_chi}}
  \end{figure*}

We use our post-processing code \textsc{\mbox{Paperboats}} to calculate the evolution of the spatial distribution of the dust material on the basis of the MHD output of \textsc{\mbox{AstroBEAR}}. We investigate the survival fraction $\eta$  for 12 lognormal grain size distributions with different peak radii $a_\text{peak}$ and a fixed distribution width $\sigma=0.1$. The survival fraction $\eta$ is defined as the ratio of total surviving dust mass (grain radii $\ge\unit[0.6]{nm}$) to the initial dust mass.

\subsubsection{Dependence on density contrast $\chi$}
\label{421}
As outlined by previous studies (\citealt{Silvia2010, Silvia2012}; \mbox{\citetalias{Kirchschlager2019b}}; \citealt{Kirchschlager2020}), the impact of sputtering and grain-grain collisions and thus the surviving dust mass depend strongly on the clump density contrast $\chi$. To start with, we compare the survival fractions $\eta$ with the results of \mbox{\citetalias{Kirchschlager2019b}}. Magnetic fields are not considered in this case, however, the novel scheme for partial grain vaporization is taken into account which is newly implemented in \textsc{\mbox{Paperboats}} (see Section~\ref{400d}). Moreover, the spatial resolution of the domain has increased by a factor of 5.

The silicate survival fraction as a function of peak radius $a_\text{peak}$ is shown for different density contrasts $\chi$ in Fig.~\ref{fig_chi} (left). This plot can be directly compared to Fig.~23 in \mbox{\citetalias{Kirchschlager2019b}}. Two significant differences are visible: Firstly, we extended the parameter space in the present study to peak radii $a_\text{peak}$ below $\unit[10]{nm}$. These grain size distributions show a substantial dust survival rate for density contrasts above $\chi=100$. Secondly, the dust survival fractions for grain sizes above a few $\unit[100]{nm}$ are much larger compared to the results in \mbox{\citetalias{Kirchschlager2019b}} which is due to the partial grain vaporization approach. 

The survival fractions are presented for density contrasts between $\chi=10$ and 1000. However, we note that density contrasts of 10 or 20 are not very realistic for the ejecta of Cas~A; $\chi$ takes rather values between ${\sim}100$ and $1000$ (e.g.,~\citealt{McKee1987, Sutherland1995, Docenko2010, Silvia2010}). In Fig.~\ref{fig_chi} we can see that the dust survival fractions for $\chi=$10 and 20 are very large (at least for $B_0=0$), only a small amount of dust is destroyed. The reason is the low gas density in the clumps which is not sufficient to sputter the dust or to accelerate the grains to appropriate velocities for catastrophic grain-grain collisions. We will constrain our analysis to the density contrasts of 50 or above, considering slightly smaller clump densities as these show a similar behaviour to $\chi=100$. 

We can then roughly distinguish between low- ($\chi \sim 50-100$) and high-density contrasts ($\chi>100$). For low-density contrasts, most of the dust mass composed of a grain size distribution with $a_\text{peak}\unit[{\sim}1]{nm}$ is destroyed. The reason is the decoupling of gas and dust due to the reduced drag in the low-density post-shock gas. The dust grains are then expelled from the clump and exposed to the hot gas of the shocked ambient medium. At temperatures around $10^8\,$K, the sputtering rates are close to their maximum  (e.g.,~\citealt{Tielens1994, Nozawa2006}) and thermal sputtering can quickly erode the grains. Increasing the grain size results in an increased dust survival fraction as the surface-to-volume ratio of the dust grains is decreasing which makes sputtering less effective. Above peak radii $a_\text{peak}=\unit[200]{nm}$, sputtering destroys less than half of the dust material. Grain-grain collisions play only a secondary role for low-density contrasts, for both  small and large grains (we will analyse below the impact of the different processes, e.g., Fig.~\ref{Fig_processes_2}). For the largest grain sizes, the survival fraction converges to 100~per~cent.

For high-density contrasts ($\chi>100$), the small grain size regime shows large survival fractions. For $\chi=100$, only 5~per~cent of the dust mass survives for a grain size distribution with $a_\text{peak}=\unit[1]{nm}$, while the survival fraction of the same grain size distribution is nearly 100~per~cent for $\chi=1000$. The higher the density contrast, the larger the preserved dust fraction. The reason for these large survival rates is the strong gas drag in the high-density gas which acts on the small dust grains. Consequently, the relative velocities between the post-shock gas and the dust grains are low and thus kinetic sputtering is not efficient. Due to the high gas densities, the pre-shock temperatures in the clumps are preserved for a long time which makes thermal sputtering also ineffective. As destruction by grain-grain collisions is more important for larger grains sizes, a large fraction of small dust grains can resist destruction. Increasing the grain size results in a reduced coupling of gas and dust and the survival fraction decreases. The increased destruction is due to sputtering at density contrasts $\chi=200$ and 300 and due to grain-grain collisions at the highest densities $\chi=1000$. The impact of sputtering decreases with increasing grain size while grain-grain collisions become more and more important. For grain size distributions with peak radii above $\unit[100]{nm}$ the dust survival fraction increases again. The reason is the partial grain vaporization which destroys in most cases only the small dust grains while most of the mass of micrometer sized grains withstands the collisions. The higher the density contrast, the lower is the survival fraction in the large particle regime.   

The total surviving dust mass (in solar masses) for a single clump is shown in Fig.~\ref{fig_chi} (right) for different density contrasts. Though the survival fraction of micrometer-sized grains is lower the higher the density contrast, the total surviving dust mass is largest for $\chi=1000$.

In the following, we will consider the density contrast $\chi=300$ as our reference case and study the impact of the magnetic field strength mainly for this density contrast.


\subsubsection{The impact of magnetic field strength}
\label{422}

   \begin{figure*}
 \resizebox{\hsize}{!}{ 
  \includegraphics[trim=2.8cm 2.1cm 0.9cm 2.0cm, clip=true,  page=1]{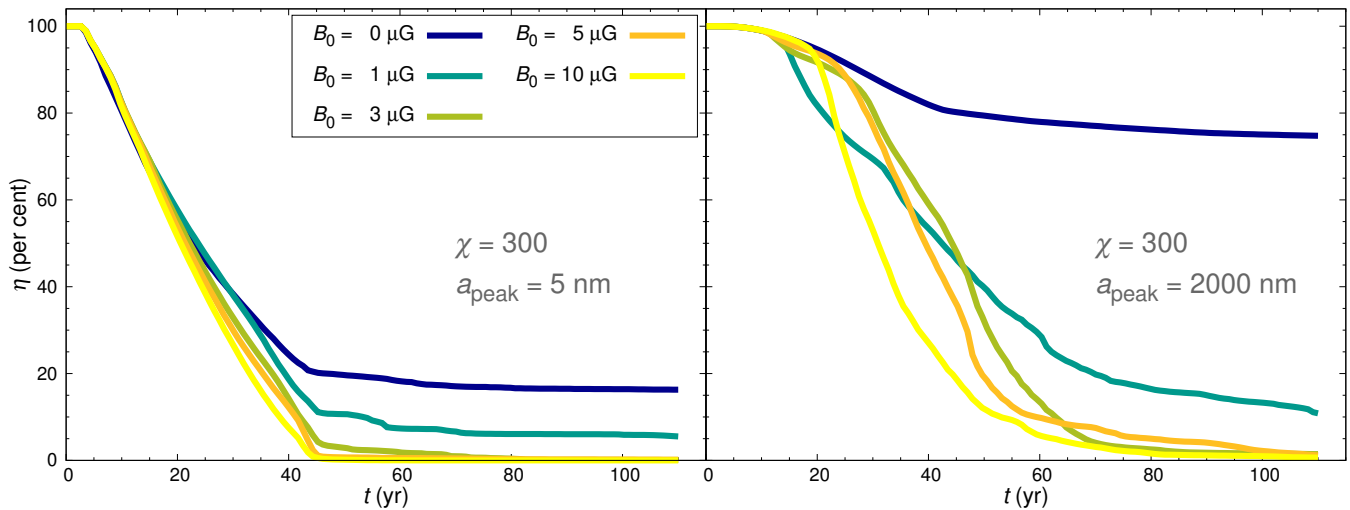}\hspace*{-0.18cm}
  \includegraphics[trim=4.85cm 2.1cm 0.9cm 2.0cm, clip=true,  page=3]{Pics/Results_paper/Survival_evolution.pdf}
  }
   \caption{Temporal evolution of the silicate survival fraction $\eta$ for different initial magnetic field strengths $B_0$. The density contrast is  $\chi=300$.  The grain sizes follow a lognormal distribution with $a_\text{peak}=5\,$nm (\textit{left}) and $2000\,$nm (\textit{right}), the distribution width is $\sigma=0.1$. For $a_\text{peak}=5\,$nm, sputtering is the main destruction process, while for $a_\text{peak}=2000\,$nm  the grains are destroyed by the combined effects of sputtering and grain-grain collisions (see Fig.~\ref{Fig_processes_2}).\label{fig_eta_B}}
   \end{figure*}

We investigate the impact of the initial magnetic field strength $B_0$ on the silicate survival fraction $\eta$. We vary the magnetic field strength between 0 and $10\,\mu$G while its pre-shock orientation is perpendicular to the shock direction.\footnote{We discuss the dust destruction for an initial magnetic field parallel to the shock direction in Section~\ref{magn_para}.} The acceleration of the grains due to the magnetic field occurs exclusively in the post-shock gas as the dust is at rest in the pre-shock gas and no relative velocity between charged grains and magnetic field exists. In the post-shock region, the compressed gas produces an amplified magnetic field (see Section~\ref{s510}) which accelerates the charged grains. 

\begin{figure*}
  \resizebox{\hsize}{!}{ 
   \includegraphics[trim=3.0cm 2.5cm 0.9cm 2.0cm, clip=true, page=1]{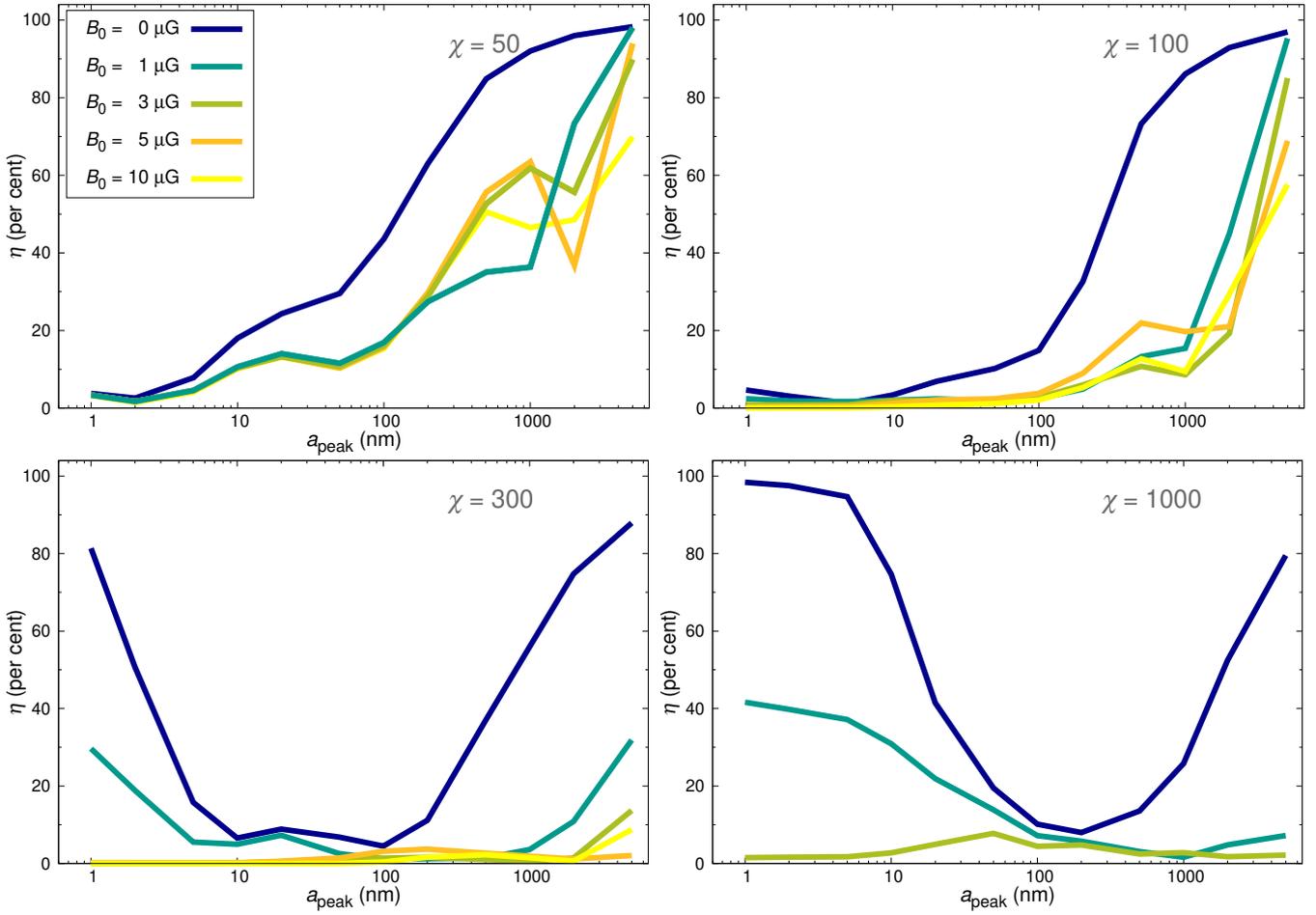}\hspace*{0.38cm} 
   \includegraphics[trim=3.0cm 2.5cm 0.9cm 2.0cm, clip=true, page=2]{Pics/Results_paper/Survival_chi_B.pdf}}\\
  \resizebox{\hsize}{!}{    
   \includegraphics[trim=3.0cm 2.5cm 0.9cm 2.0cm, clip=true, page=3]{Pics/Results_paper/Survival_chi_B.pdf}\hspace*{0.38cm} 
   \includegraphics[trim=3.0cm 2.5cm 0.9cm 2.0cm, clip=true, page=4]{Pics/Results_paper/Survival_chi_B.pdf}   
   }
   \caption{Silicate survival fraction $\eta$ for different density contrasts  $\chi$, magnetic field strengths $B_0$, and peak radii $a_\text{peak}$. The MHD simulations for $B_0=5\,\mu$G and $10\,\mu$G are very time-consuming and were not feasible in a reasonable amount of time for the highest density contrast $\chi=1000$. However, based on lower magnetic field strengths we expect nearly complete destruction for all dust grains sizes at these large magnetic fields strengths.
   \label{Fig_chi_B2}}
   \end{figure*}

In Fig.~\ref{fig_eta_B} the temporal evolution of the dust survival fraction is shown for the density contrast $\chi=300$ and for grain size distributions with $a_\text{peak}=\unit[5]{nm}$ and $a_\text{peak}=\unit[2000]{nm}$. We choose to show these grain sizes as their destruction is dominated by different processes: The $\unit[5]{nm}$ grains are mainly destroyed by sputtering. The $\unit[2000]{nm}$ grains are destroyed by the combined effects of sputtering and grain-grain collisions, which effect prevails depends on the initial magnetic field strengths (see Fig.~\ref{Fig_processes_2}). Following Fig.~\ref{fig_eta_B}, it is obvious that magnetic fields $B_0>0$ cause a larger dust destruction compared to the case without magnetic field. We will outline the reasons for this behaviour in Section~\ref{sec4_diffdustprocesses}. The rough trend is that the dust survival fraction decreases with increasing magnetic field strength, though exceptions exist, in particular for micrometre sized grains. This is in agreement with Fig.~\ref{Fig_chi_B2} where the dust survival fraction after three cloud-crushing times is shown for different density contrasts $\chi$ and peak radii $a_\text{peak}$. The data of the non-magnetic case at the density contrast $\chi = 300$ are taken from Fig.~\ref{fig_chi}. Similar to the previous cases we see that the dust destruction is larger for higher magnetic field strengths. Only for grain sizes above a few ${\sim}100\,$nm is no clear trend visible though the largest dust survival fraction for all density contrast is for $B_0=0$. 

In general, the survival fractions under the influence of magnetic fields show similar characteristics as those without magnetic fields, however, at lower levels. Depending on the initial grain sizes and the density contrast, the survival fraction at a magnetic field strength of $1\,\mu$G can decrease between nearly zero and 60 per cent compared to the non-magnetic case.

 \begin{figure*}
  \resizebox{\hsize}{!}{ 
   \includegraphics[trim=3.0cm 2.5cm 0.9cm 2.0cm, clip=true, page=3]{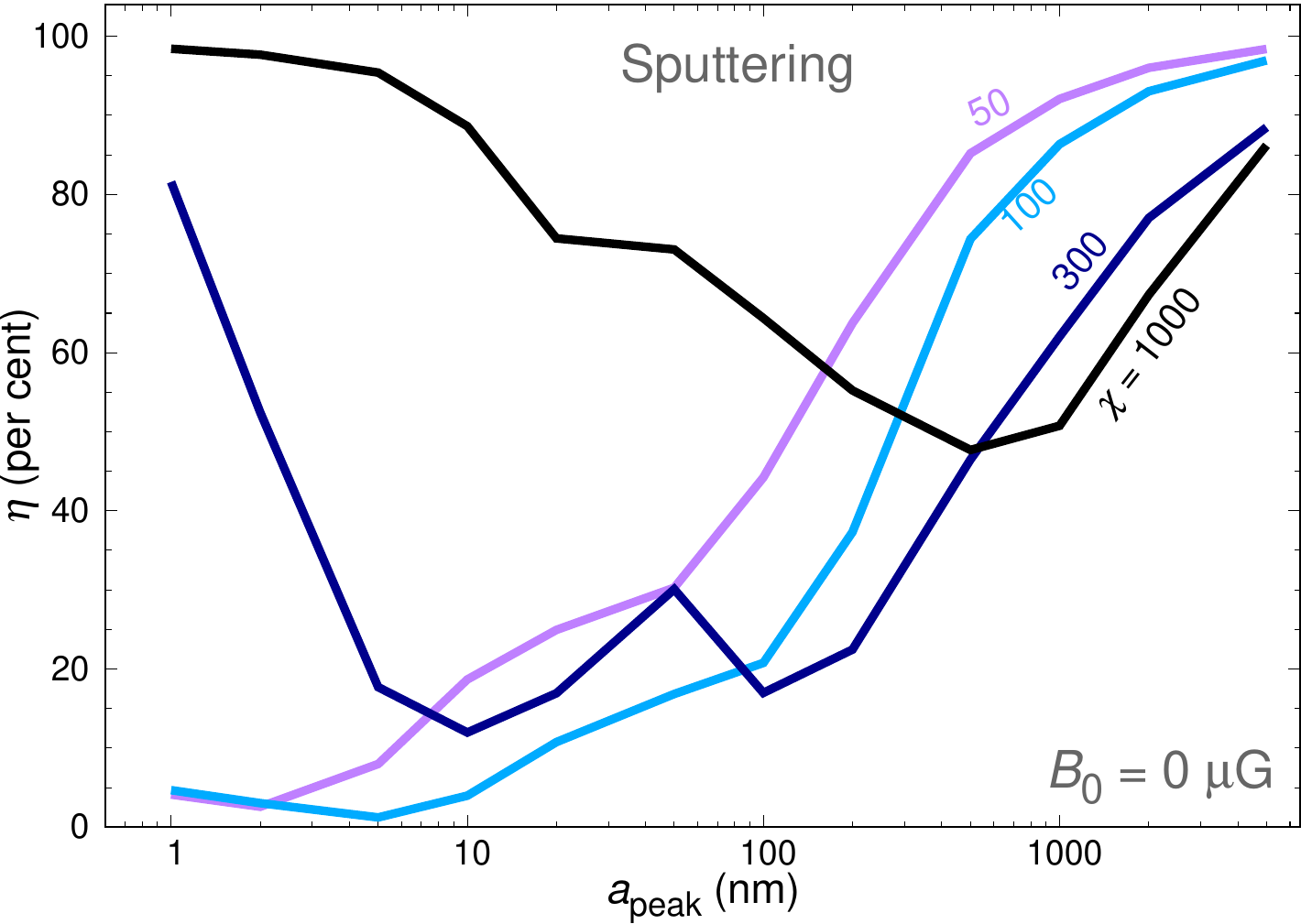}\hspace*{0.38cm} 
   \includegraphics[trim=3.0cm 2.5cm 0.9cm 2.0cm, clip=true, page=2]{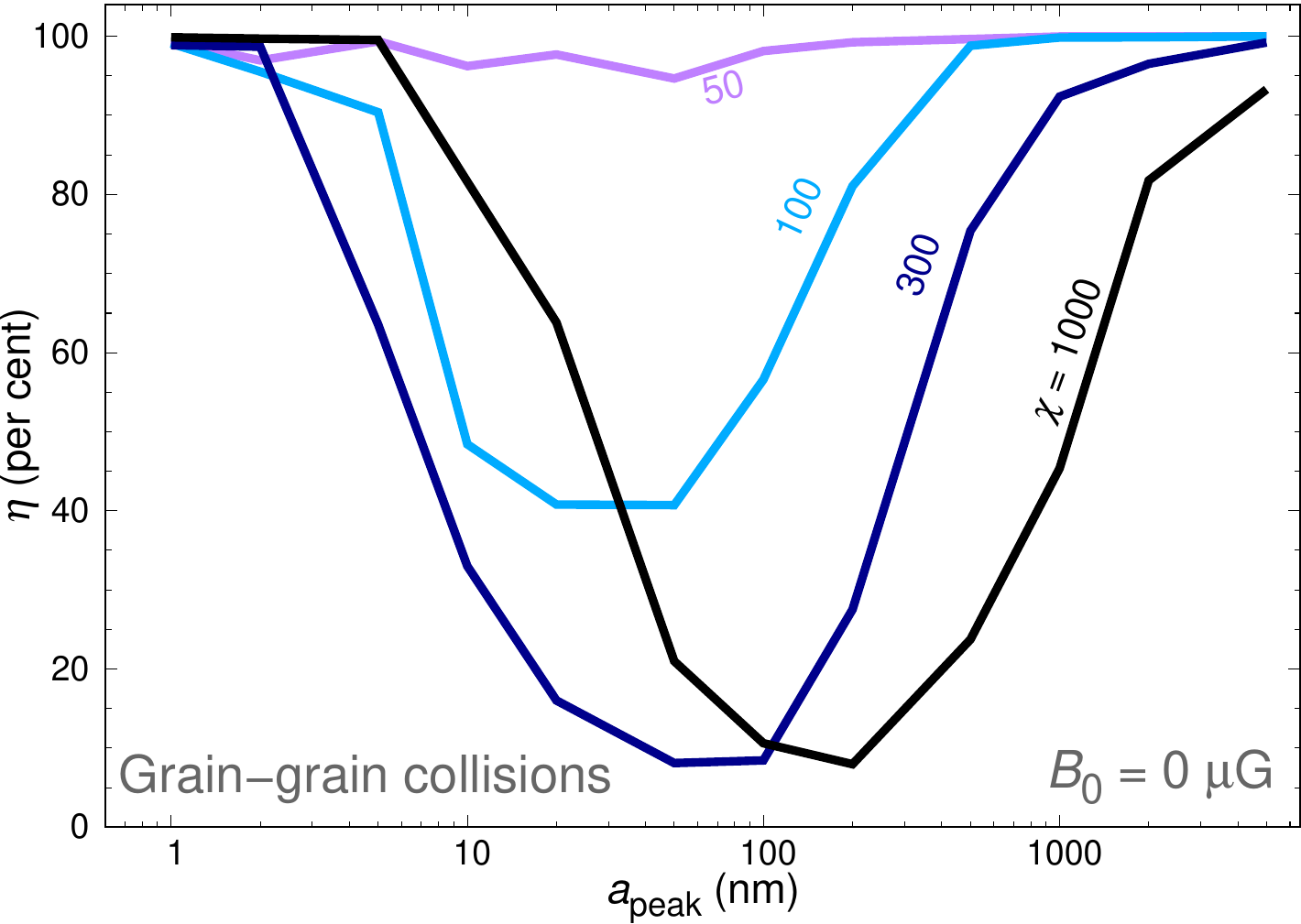}}\\
  \resizebox{\hsize}{!}{    
   \includegraphics[trim=3.0cm 2.5cm 0.9cm 2.0cm, clip=true, page=5]{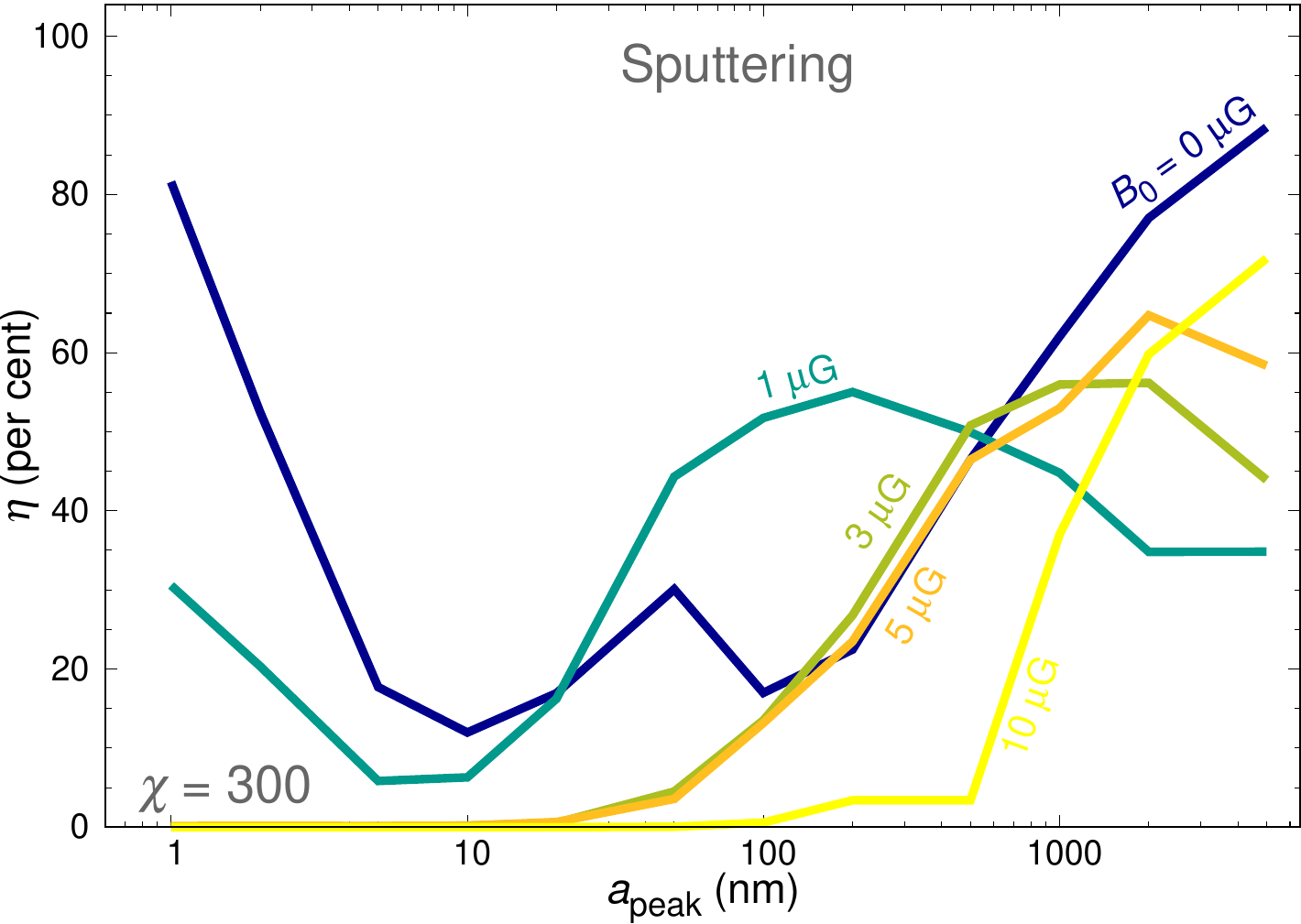}\hspace*{0.38cm} 
   \includegraphics[trim=3.0cm 2.5cm 0.9cm 2.0cm, clip=true, page=4]{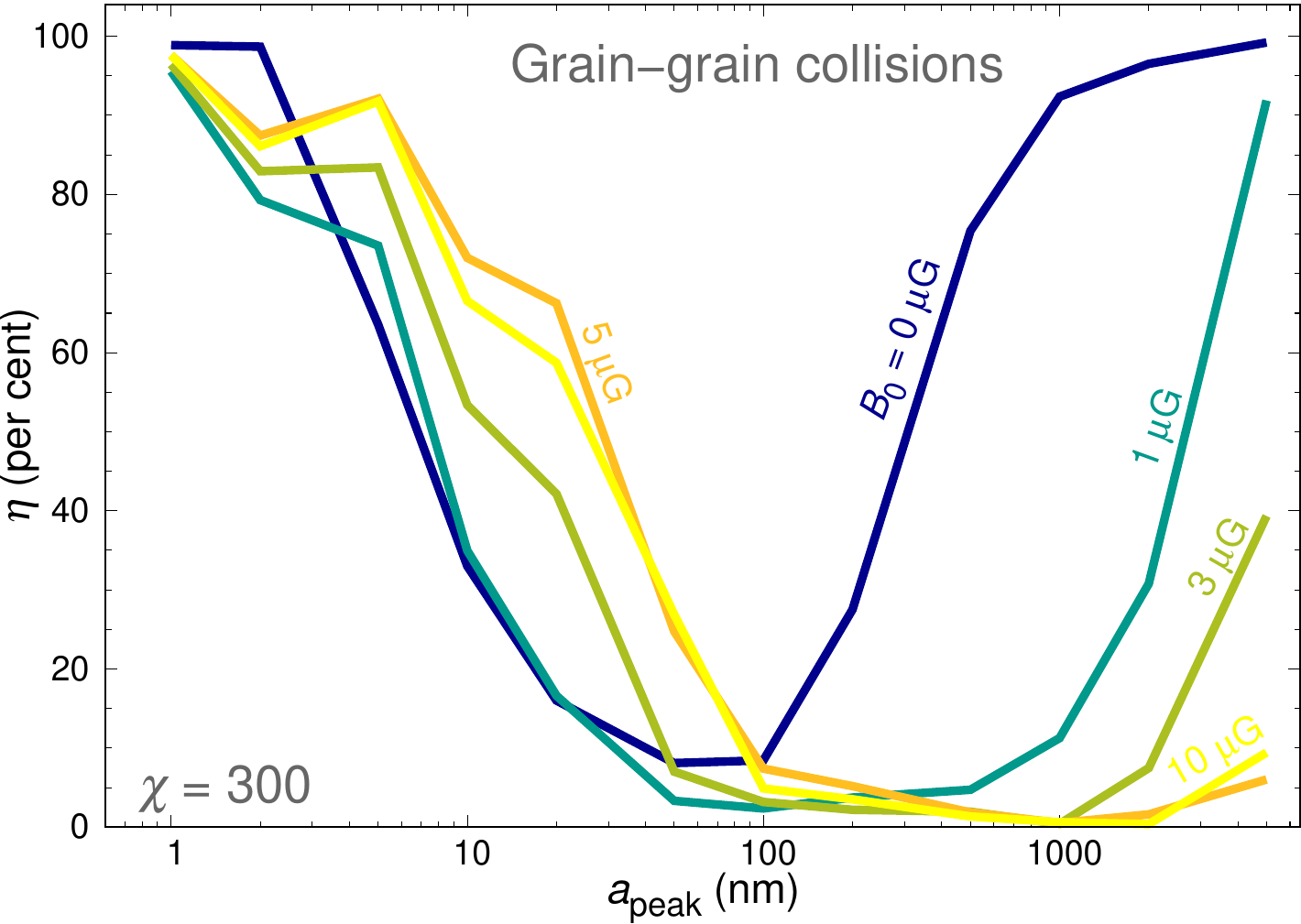}   
   }
   \caption{Silicate survival fraction $\eta$ as a function of the initial grain size $a_\text{peak}$ for different processes: Sputtering only (\textit{left}) and grain-grain collisions only (\textit{right}). In the \textit{top} \textit{row}, the density contrast $\chi$ is varied and the initial magnetic field strength is fixed to $B_0=0\,\mu$G. In the \textit{bottom} \textit{row}, the  magnetic field strength $B_0$ is varied and the density contrast is fixed to $\chi=300$.\label{Fig_processes_2}}
   \end{figure*}  
  
  \begin{figure*}
  \resizebox{\hsize}{!}{ 
   \includegraphics[trim=3.0cm 2.5cm 0.9cm 2.0cm, clip=true, page=1]{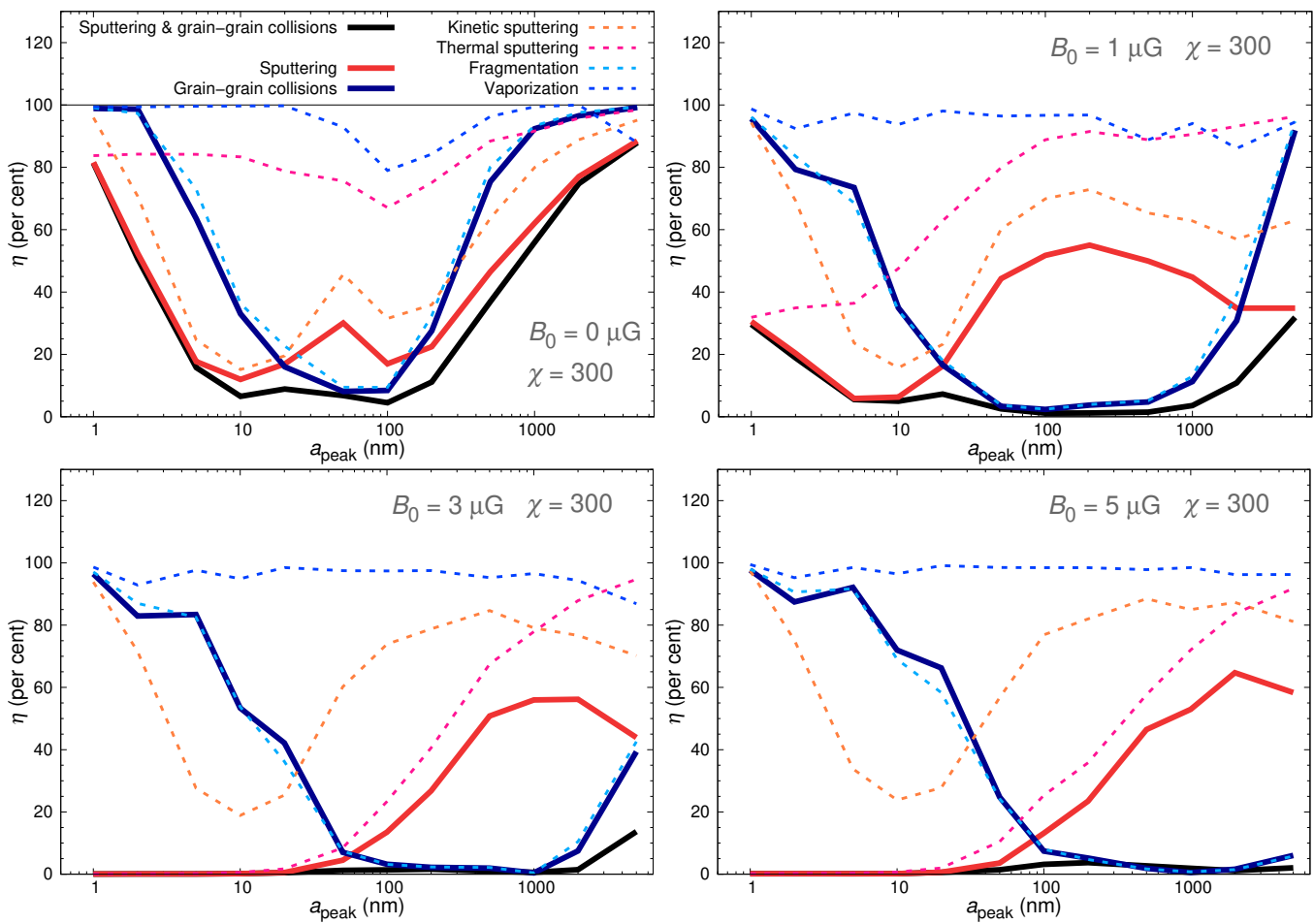}\hspace*{0.38cm} 
   \includegraphics[trim=3.0cm 2.5cm 0.9cm 2.0cm, clip=true, page=2]{Pics/Results_paper/Survival_processes_all.pdf}}\\
  \resizebox{\hsize}{!}{    
   \includegraphics[trim=3.0cm 2.5cm 0.9cm 2.0cm, clip=true, page=3]{Pics/Results_paper/Survival_processes_all.pdf}\hspace*{0.38cm} 
   \includegraphics[trim=3.0cm 2.5cm 0.9cm 2.0cm, clip=true, page=4]{Pics/Results_paper/Survival_processes_all.pdf}   
   }
    \caption{Silicate survival fraction $\eta$ for different processes: Sputtering plus grain-grain collisions (black), sputtering only (red), grain-grain collisions only (blue), kinetic sputtering only (orange), thermal sputtering only (pink), fragmentation only (light green), vaporization only (light blue). The initial magnetic field strength $B_0$ is varied, the density contrast is fixed to $\chi=300$.\label{Fig_processes_all}}
   \end{figure*}  
  
   \begin{figure}
  \resizebox{\hsize}{!}{ 
   \includegraphics[trim=3.0cm 2.5cm 0.9cm 2.0cm, clip=true, page=1]{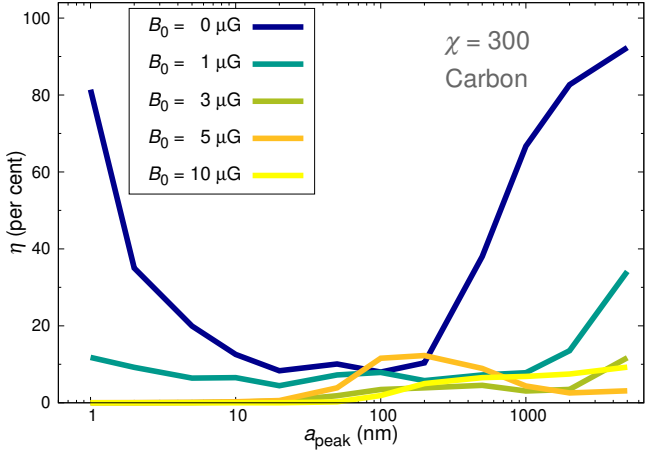}
   }
   \caption{Carbon survival fraction $\eta$ for different magnetic field strengths $B_0$ and peak radii $a_\text{peak}$. The density contrast is $\chi=300$ (compare with Fig.~\ref{Fig_chi_B2}, \textit{bottom} \textit{left}).\label{fig_compo}}
   \end{figure} 
     \begin{figure}
  \resizebox{\hsize}{!}{ 
   \includegraphics[trim=3.0cm 2.5cm 0.9cm 2.0cm, clip=true, page=1]{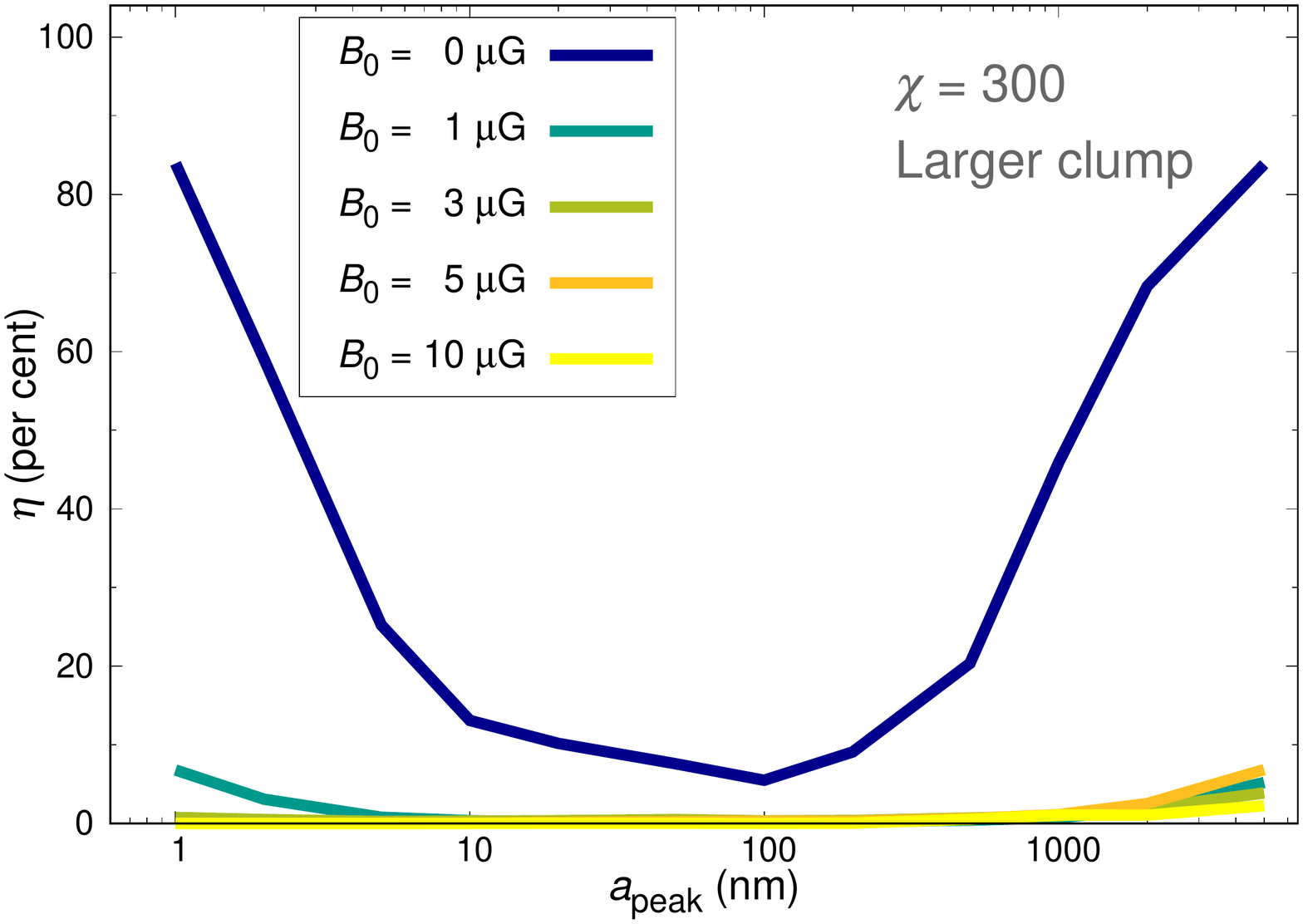}
   }
   \caption{Silicate survival fraction $\eta$ in a clump with its diameter doubled to $4\times10^{16}\,$cm (compare with Fig.~\ref{Fig_chi_B2}, \textit{bottom} \textit{left}).\label{fig_clumpsize}}
   \end{figure}
  
\subsubsection{The different dust destruction processes}
\label{sec4_diffdustprocesses}
 
In this section we investigate the impact of the density contrast and the magnetic field strength on the individual destruction processes, namely sputtering and grain-grain collisions. Therefore, we performed simulations with \textsc{\mbox{Paperboats}} in which either sputtering or grain-grain collisions are turned off and determined the destroyed dust masses. Please note that simulations with sputtering only, with grain-grain collisions only, or with both processes, are three different simulations. In general, the amount of dust destroyed by the combined effects is different to the sum of dust destroyed by the individual processes. This synergistic effect was already identified in \citetalias{Kirchschlager2019b}.

In Fig.~\ref{Fig_processes_2} the dust survival fraction $\eta$ of the two processes is shown for different density contrasts $\chi$ and fixed magnetic field strength $B_0=0\,\mu$G (\textit{top} \textit{row}), as well as for different magnetic field strengths $B_0$ and fixed density contrast $\chi=300$ (\textit{tbottom} \textit{row}).

Following the discussion in Section~\ref{421}, we see in the \textit{top} \textit{row} of Fig.~\ref{Fig_processes_2} that the density contrast has an impact on both the sputtering and the grain-grain collisions. At low-density contrasts ($\chi \sim 50-100$), the destruction is dominated by sputtering for all grain sizes. Grain-grain collisions play only a secondary role as the dust number densities and thus the collision probabilities are too small (see equation~(22) in \mbox{\citetalias{Kirchschlager2019b}}). Destruction by sputtering decreases nearly monotonically with increasing grain size. For high-density contrasts ($\chi>100$), this behaviour is broken by an increased survival fraction in the small grain size range due to the better coupling between gas and dust. Grain-grain collisions become dominant at high-density contrasts in the medium grain size regime (between ${\sim}10\,$nm and a few $100\,$nm) where significant destruction occurs. Small dust grains experience less destruction by grain-grain collisions as they are well coupled to the gas which reduces the relative velocities between the grains. At the other end of the grain size range, micrometre grains are better preserved as most of the collisions occur between small and large grains which show the largest relative velocities: For a collision of dust grains with a significant difference in size, only a small amount of the dust mass is destroyed due to partial vaporization, in most cases approximately twice the mass of the smaller dust grain, unless the collision velocity is extremely high and can vaporize also a substantial fraction of the big grain.

Following the discussion in Section~\ref{422}, we see in the \textit{bottom} \textit{row} of Fig.~\ref{Fig_processes_2} the destruction by sputtering and grain-grain collisions as a function of the magnetic field strength. The charged grains get an additional acceleration which increases the relative velocities between gas and dust as well between dust grains of different sizes. Moreover, the number of grain-grain collisions and gas-grain collisions is increased. This has different implications for the efficiency of sputtering and grain-grain collisions. For the sputtering, the main impact of the magnetic field is on the small grain regime. After shock compression, the grains are expelled from the high density regions in the post-shock gas and are more easily sputtered in the hot gas of the ambient medium. The effect is larger the larger the magnetic field strength, and most of the $10\,$nm grains or smaller are destroyed for $B_0\ge3\,\mu$G. Furthermore, the higher relative velocities between gas and dust cause also a larger destruction by sputtering at other grain sizes, but the decreased surface-to-volume ratio of the medium and micrometre sized grains prevents total destruction by sputtering. For the grain-grain collisions, on the other hand, magnetic fields have no significant effect on the small grains ($\lesssim10\,$nm) as the collision probabilities are too low. At medium grain sizes and micrometre grains, the large grain velocities induced by the magnetic field acceleration enable high grain-grain collision velocities and thus a higher destruction efficiency. These collision velocities are even high enough to overcome the partial vaporization which protects a large fraction of dust mass in the micrometre grains for $B_0=0$, but which nearly results in total destruction for $B_0\ge5\mu$G ($\eta<2\,$per~cent for $a_\text{peak}=1000\,$nm). 


In Fig.~\ref{Fig_processes_all}, the dust destruction processes are studied to an even deeper level: sputtering is split into kinetic and thermal sputtering, grain-grain collisions are split into fragmentation and vaporization. For each process we run a separate simulation with \mbox{\textsc{Paperboats}} and switch other destruction processes off. The magnetic field strength is varied while the density contrast is fixed to $\chi=300$. The different survival fractions for kinetic and thermal sputtering support the discussion above. With increasing magnetic field strength, small dust grains are better coupled to the gas. As a consequence, the destruction by thermal sputtering is becoming more and more important. The interpretation of the simulations for vaporization and fragmentation is more complex. Vaporization without fragmentation or other destruction processes destroys only a small amount of dust material. The reason is that the initial grain size distributions are narrow and vaporization does not lead to a significant re-distribution of the dust in the grain size bins. As a result, most of the dust grains have similar sizes and thus low relative velocities, resulting in only a minimum of dust destruction by vaporization. On the other hand, if vaporization is combined with a process that is able to re-distribute dust grains over a wide grain size range, as fragmentation does, the dust destruction efficiency is much higher.

In summary, we see that the destruction of dust grains in the clumpy ejecta is a complex interplay between the four destruction processes (kinetic and thermal sputtering, fragmentation, vaporization) and the dust transport due to gas and plasma drag plus magnetic field acceleration: The destruction processes influence the dust transport and vice versa.

\subsubsection{Carbon dust}

The results presented so far have been for silicate dust which is consistent with the very oxygen-rich composition of the ejecta of Cas~A (\citealt{Chevalier1979}). However, carbon grains may be an important dust component in many SNRs and we will briefly compare the destruction of carbon grains in a Cas~A ejecta clump with that of silicate. 

Fig.~\ref{fig_compo} shows the carbon survival fraction $\eta$ for different magnetic field strengths. The density contrast is fixed to $\chi=300$. The results can be directly compared to the silicate case (Fig.~\ref{Fig_chi_B2}, \textit{bottom} \textit{left})). Both materials show the same  qualitative behaviour. For $B_0=0$, the $1\,$nm grains are destroyed by ${\sim}20\,$per~cent. At medium grain sizes (${\sim}10-100\,$nm), the dust survival fraction has its minimum ($5-10\,$per~cent), while it rises again up to $\eta\sim90\,$per~cent for micrometre grains. For $B_0>0$, the simulations yield for both materials $\eta\lesssim35\,$per~cent. 

The main difference between the dust compositions is an increase of the survival fraction for carbon grains up to $\eta\sim12\,$per~cent around $100\,$nm at $B_0=5\,\mu$G. For silicate grains, an increased survival fraction is visible at the same magnetic field strength and grain sizes, but at a lower level ($4\,$per~cent). The differences are minor and go back to a different efficiency of sputtering and grain-grain collisions as well as a different charging of silicates and carbon, but also to a different bulk density of the dust materials which affects the gas drag and number densities.


\subsubsection{Clump size}
\label{sec_cs}
The structure of Cas~A is highly clumped (\citealt{Fesen2006, Milisavljevic2013}) and the clump sizes seen in observations are in the range $(1-5)\times10^{16}\,$cm (diameter; \citealt{Fesen2011}). We can investigate whether the clump size has a significant impact on the survival fraction $\eta$. Therefore, we doubled the clump diameter in our setup from $2\times10^{16}\,$cm to $4\times10^{16}\,$cm, ran MHD~simulations using \textsc{\mbox{AstroBEAR}} for the large clump, and subsequently  performed post-processing simulations using \textsc{\mbox{Paperboats}} for silicate dust. The results are shown in Fig.~\ref{fig_clumpsize} and can be directly compared to the case of the smaller clump size (Fig.~\ref{Fig_chi_B2}, \textit{bottom} \textit{left})).

For $B_0=0$, the survival fraction $\eta$ in the small grain size regime ($<20\,$nm) is increased by $5-10\,$per~cent for the large clump. In the shocked large clump it is harder for the grains to be ejected into the hot ambient medium. On the other hand, the survival fraction of large grains ($>200\,$nm) is decreased by $5-10\,$per~cent for the large clump. The shocked large clump shows higher gas densities compared to the small clump which increases the grain number densities and thus the destruction efficiency by grain-grain collisions. For \mbox{$B_0>0$}, only the magnetic field with $B_0=1\,\mu$G shows a significant survival fraction in the small grain size range ($\eta{\sim}7\,$per~cent at $a_\text{peak}=1\,$nm). Larger magnetic fields result in the destruction of more than $99\,$per~cent of the initial dust mass in the large clump for all grain sizes, only micrometre sized grains show a survival fraction up to $\eta\sim7\,$per~cent due to partial vaporization.

In summary, the destruction is at a similar level for the two clump sizes for the density contrast $\chi=300$. We expect that larger differences can be detected with increasing density contrast. As grain-grain collisions are becoming more and more important for high gas densities, this will result in larger dust survival fractions for small clumps compared to large clumps. In all cases, a magnetic field strength $B_0>0$ reduces the survival fractions even further.

  \subsubsection{Dust destruction in a magnetic field parallel to the shock direction} 
 \label{magn_para}
        \begin{figure}
 \resizebox{\hsize}{!}{ 
  \includegraphics[trim=2.8cm 2.5cm 0.9cm 2.0cm,  clip=true,  page=1]{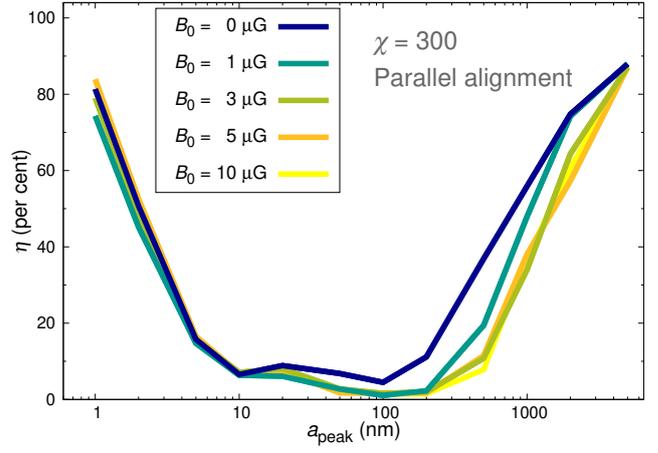}
  }
  \caption{Silicate survival fraction $\eta$ as a function of the initial magnetic field strength $B_0$ for an initial magnetic field orientation parallel to the shock direction. The density contrast is $\chi=300$. The influence of the magnetic field is small at this orientation.\label{fig_parallel}}
  \end{figure}
In the previous sections the dust survival fraction for magnetic field orientations perpendicular to the shock propagation was studied. Here, we want to analyse the second extreme case of magnetic field orientation when the initial magnetic field lines are orientated parallel to the shock direction. We expect a less effective destruction of the dust grains as the gas flow is less disrupted by the field lines and the grains' main direction of motion is parallel to the magnetic field.

In Fig.~\ref{fig_parallel}, the dust survival fractions $\eta$ for different magnetic field strengths $B_0$ are shown. The density contrast is $\chi=300$. Obviously, the parallel magnetic field has nearly no influence on the survival fraction of grains smaller than $10\,$nm. However, for grains larger than $10\,$nm, the survival fraction slightly decreases with increasing magnetic field strength, though to a significantly lower degree compared to the perpendicular case. For $a_\text{peak}=1000\,$nm, for example, the survival fraction at $B_0=10\,\mu$G is $15\,$per~cent below that for $B_0=0$.  The acceleration of the dust grains due to the magnetic field (equation~\ref{Lorentz}) amounts to zero when the shock hits the clump and only contributes to the dust motion if the magnetic field morphology is disturbed by the shocked gas.

%




 \section{Discussion}
 \label{500}
 \subsection{Grain size survival scheme}
 \label{510}
    \begin{figure}
  \resizebox{\hsize}{!}{ 
   \includegraphics[trim=2.2cm 2.1cm 1.6cm 2.2cm, clip=true, page=1
   ]{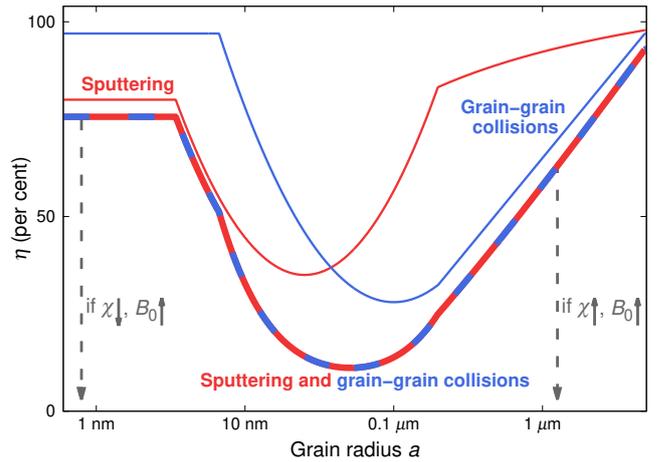}
   }
   \caption{Schematic overview of the efficiency of sputtering (red), grain-grain collisions (blue), or the combined effects of sputtering and grain-grain collisions (red-blue dashed) at various clump densities and magnetic field strengths. For grain survival fractions $\eta$ versus grain radius $a$, three different grain size regions can be identified:  Small grains with radii ${\lesssim}10\,$nm, medium sized grains with radii between ${\sim}10$ and a few $100\,$nm, and larger (sub-)micrometre sized grains (${\gtrsim}100\,$nm).\label{fig_scheme_processes}}
   \end{figure} 
       \begin{figure*}
 \resizebox{\hsize}{!}{ 
  \includegraphics[trim=2.8cm 2.4cm 0.9cm 2.0cm, clip=true,  page=1]{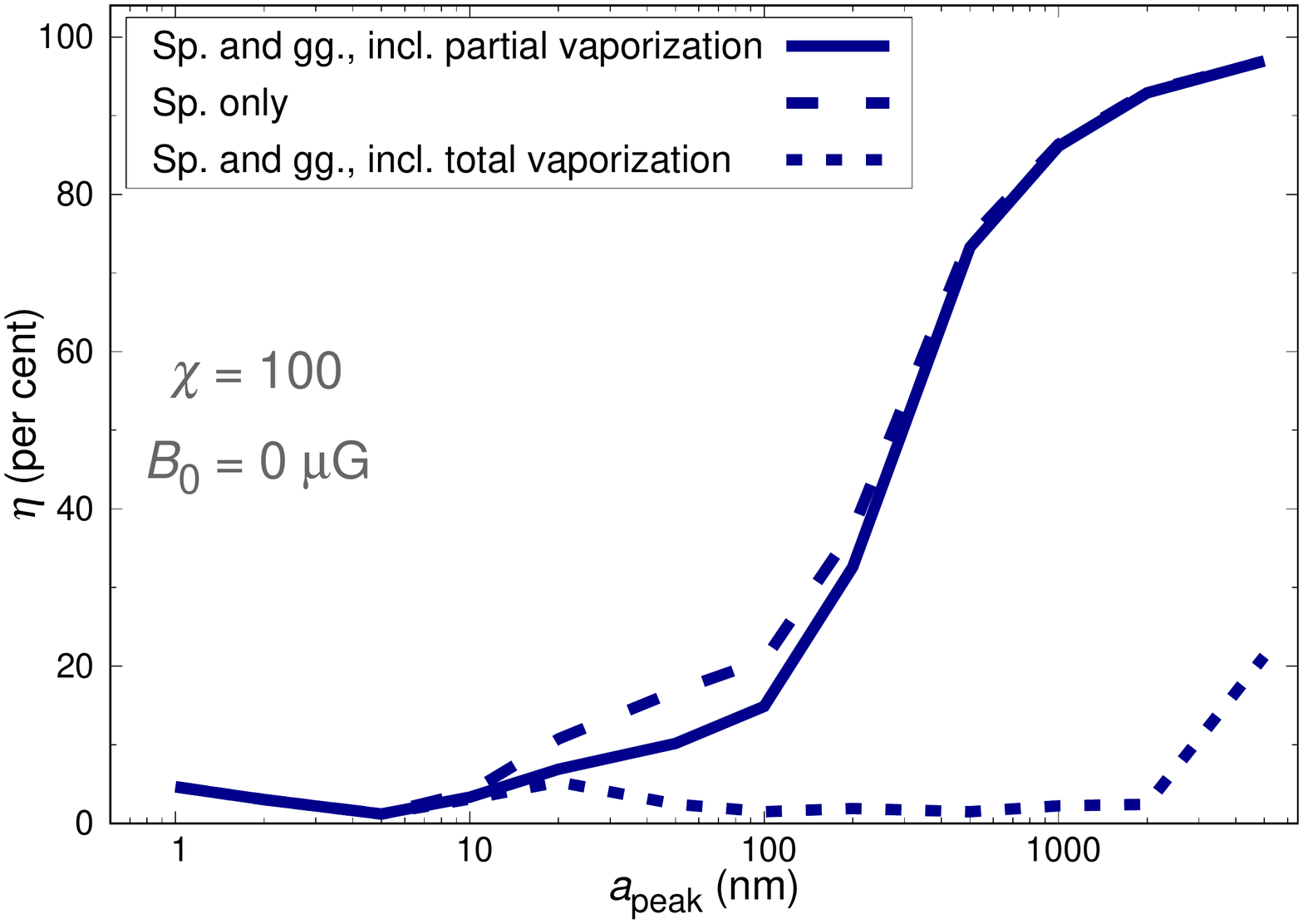}
  \includegraphics[trim=1.5cm 2.4cm 0.9cm 2.0cm, clip=true,  page=2]{Pics/Results_paper/Impact_of_gg.pdf}
  }
  \caption{Comparison of the silicate survival fraction of different approaches for the grain-grain collisions: Destruction by sputtering (Sp.) and grain-grain collisions (gg.), including partial vaporization (solid line; new approach); destruction by sputtering only (dashed line; no grain-grain collisions); destruction by  by sputtering and grain-grain collisions, including total vaporization (dotted line; approach of \citetalias{Kirchschlager2019b}).\label{fig_gg_old_and_new}}
  \end{figure*}  
   
  Based on the analysis of the previous section, we can sketch an overview that shows schematically which grain sizes are able to survive the passage of the reverse shock and which processes are the dominant factor. The scheme in Fig.~\ref{fig_scheme_processes} shows the survival fraction for grain-grain collisions, sputtering, and the combined effects of grain-grain collisions and sputtering for different grain radii $a$. The plots roughly approximate the behaviour for the density contrast \mbox{$\chi=300$} at $B_0=0$ though we try to extrapolate to other density contrasts and magnetic field strengths through arrows indicating an decrease of the survival fraction when $\chi$ or $B_0$ increase or decrease.
  
  We identify three grain size regimes which show different behaviours when it comes to avoiding dust destruction: Small grains with radii ${\lesssim}10\,$nm, medium sized grains with radii between ${\sim}10$ and a few $100\,$nm, and larger (sub-)micrometre sized grains (${\gtrsim}100\,$nm).
  
  \begin{itemize}
   \item  Small grains (${\lesssim}10\,$nm) can have high survival fractions for high-density contrasts and low magnetic field strengths as the gas drag is very efficient in this case and the dust grains are better protected in dense fragments of the disrupted clump. The survival fraction decreases for low-density contrasts as this causes the decoupling of gas and dust and thus less protection. As can be seen in Fig.~\ref{Fig_processes_all}, it is the increased thermal sputtering and not the kinetic sputtering that destroys large amounts of small dust grains when the magnetic field strength rises. Grain-grain collisions can be ignored for these grains.
   \item  On the other side of the size range, (sub-)micrometre grains (${\gtrsim}100\,$nm) are able to survive the reverse shock to a high degree for small density contrasts and low magnetic field strengths. The efficiency of sputtering decreases with increasing grain size because of the reduced surface-to-volume ratio. The efficiency of grain-grain collisions is also decreasing with increasing grain size as the collision velocities are able to fragment and vaporize only a small part of the dust grains. The survival fraction decreases for large density contrasts or higher magnetic field strengths as this causes a larger dust number density and higher collision velocities, which results in more frequent and more catastrophic grain-grain collisions. 
   \item The medium grain sizes (${\sim}10$ to a few $100\,$nm) experience a mixture of the effects experienced by the small and  the micrometre grains. For low-density contrasts, sputtering is the main effect acting on the medium grain sizes. A significant amount of dust can survive for $B_0=0$ but magnetic fields reduce the surviving dust mass. For high-density contrasts, both sputtering and grain-grain collisions are efficient and can destroy mostly all of the dust material, with or without magnetic field. Nevertheless, as can be seen in Figs.~\ref{Fig_chi_B2} and~\ref{Fig_processes_all}, there is a grain size in that range at which the dominant dust destruction process (sputtering or grain-grain collisions) changes, representing a small peak in the survival fraction, though at a low level. This peak was already identified in \citetalias{Kirchschlager2019b} for both silicate and carbon grains. For the density contrast $\chi=300$, we see that the peak shifts to larger grain sizes with increasing magnetic field, from $a_\text{peak}=20\,$nm at $B_0 = 0$ to $200\,$nm at $B_0 = 10\,\mu$G. However, the survival fraction at this peak decreases at the same time, from $\eta=6.5\,$per~cent at $B_0 = 0$ to $2.3\,$per~cent at $B_0 = 10\,\mu$G.
  \end{itemize}
To summarize, large survival fractions up to $100\,$per~cent are expected for small grains (${\lesssim}10\,$nm) if the density contrast is large, or for (sub-)micrometre grains (${\gtrsim}100\,$nm) at predominately low-density contrasts. In both cases, initial magnetic field strengths of a few $\mu$G are able to destroy huge amounts of these dust grains. For the medium grain size range, large survival fractions (but significantly lower than $100\,$per~cent) are predicted for low-density contrasts only.

 \subsection{Importance of grain-grain collisions}
 \label{520}

Besides the implementation of magnetic fields effects, one of the main improvements of \textsc{\mbox{Paperboats}} is the new approach for the partial grain vaporization. In \citetalias{Kirchschlager2019b}, we outlined the importance of grain-grain collisions. In particular for large grains and broad initial size distributions, grain-grain collisions can destroy most of the dust material. However, with the new approach for the partial grain vaporization, the largest grains are better preserved. We check now whether grain-grain collisions are still crucial for the dust destruction computations in SNRs. 

In Fig.~\ref{fig_gg_old_and_new} we compare the dust survival fractions for the approach of \citetalias{Kirchschlager2019b} and the new approach (no magnetic fields). For both presented density contrasts ($\chi=100$ and 1000), the differences are clearly visible and are the largest for grain sizes above $100\,$nm. In agreement with Section~\ref{400}, the dust survival fraction with the new approach is much larger for (sub-)micrometre grains. For the low-density contrast ($\chi=100$), the destruction occurs mostly by sputtering, and the difference between destruction by sputtering only and by grain-grain-collisions (with the partial vaporization approach) is less than 5~per~cent. On the other hand, for the high-density contrast ($\chi=1000$), the grain-grain-collisions (with the approach of partial vaporization) destroy up to 60~per~cent more dust compared to the case without grain-grain collisions. Therefore, destruction by grain-grain collisions is still important and can not be neglected. 

 \subsection{Extrapolation from the cloud-crushing problem to the full remnant}
 \label{530} 
In our study we applied the cloud-crushing problem and considered only a single clump that is impacted by a shock wave instead of modelling the entire remnant
in which the shock impacts the ejecta material and the embedded overdense gas and dust clumps. This allows us to follow the clump destruction at much higher resolution.
Assuming that the reverse shock impacts all ejecta clumps in a similar way, the results of the cloud-crushing problem can be extrapolated to the entire remnant.

However, the approach misses the global evolution of the ejecta remnant. The expansion of the remnant causes a decrease of the ejecta density (see, e.g., \citealt{Micelotta2016}) and thus a change of density contrasts. Moreover, the relative velocity difference between reverse shock and ejecta drops with time which reduces the impact velocity of the shock in the clump. Because of Cas~A's asymmetric structures and the three-dimensional morphology, the shock velocity will not be uniform at a certain time and can also point in different directions (\citealt{Vink2022a}).  It is even unclear whether each ejecta clump will be hit by the reverse shock or if some are able to survive the ejecta phase. Furthermore, the location of the clumps regarding the reverse shock wave determines the time at which they can be hit by the shock and sets the surrounding gas conditions as the gas density, gas temperature, and magnetic field strengths and direction. The shape, size and distribution of the clumps can disturb the shock front and thus alter the disruption of the clumps. 

All these effects have an influence on the total amount of dust that can survive the energetic shocks in Cas~A and the survival fractions derived from the cloud-crushing problem can either over- or underestimate the dust survival rate. A full consideration of these effects is complex and  beyond the scope of the paper and must be postponed to the future.

 \subsection{Comparison to previous studies}
 \label{540} 
 In Section~\ref{600}, we mentioned a few studies that considered or discussed the impact of magnetic fields on charged grains in dust destruction environments. In this section we want to briefly outline results of those studies which are comparable to our work here. An extensive comparison for the non-magnetic field case has been made already in Section~6 of \citetalias{Kirchschlager2019b}.

\cite{Hu2019} used 3D hydrodynamical simulations to investigate dust destruction via sputtering. They ignore magnetic fields but consider betatron acceleration of dust grains under the assumption of flux-freezing. Moreover, they study the destruction of pre-existing dust in the multiphase ISM impacted by a SN blast wave. In this scenario, the gas densities, gas temperature, gas composition, shock velocities, dust number densities and grain sizes are completely different compared to the dusty clumps in the ejecta of Cas~A. For example, the average gas number density in the diffuse ISM is of the order ${\sim}0.1-1\,$cm${}^{-3}$ and in the Cas~A ejecta clumps ${\sim}100-1000\,$cm${}^{-3}$, which is a crucial factor for the dust survival rates. Depending on the grain size and the shock velocity, the higher gas densities can either protect the grains or lead to a higher destruction rate. 
Moreover, the initial dust grains in the model of \cite{Hu2019} follow a standard MRN distribution (\citealt{Mathis1977}) which is applicable for
Milky Way-like dust in the diffuse ISM. In contrast, we consider lognormal size distributions in the SNR ejecta, which are narrower than the MRN distribution. In \citetalias{Kirchschlager2019b} we could show that the survival rates significantly depend on the initial grain size distribution (power-law or lognormal) and in particular on the width of the initial distribution.
To summarize, a comparison of the results from \cite{Hu2019} and our work is difficult. We further note that \cite{Hu2019} include sputtering as destruction process but not grain-grain collisions which would in any case yield higher survival rates.

Similarly to \cite{Hu2019}, \cite{Slavin2004, Slavin2015} and \cite{Bocchio2014}  study the destruction of pre-existing dust in the ISM by interstellar shocks which makes a comparison to our work difficult for the same reasons. The magnetic field in these studies is frozen to the gas and only the perpendicular component is considered. We note that \cite{Slavin2004, Slavin2015} and \cite{Bocchio2014} consider sputtering, fragmentation and vaporization as destruction processes.

The only study we are aware of that considers magnetic fields and destruction of grains formed in the ejecta of a SNR is by \cite{Fry2020}. However, they follow the injection of these dust grains into the ISM environment of the SNR. They assumed a turbulent magnetic field in the ISM but no magnetic field in the SNR ejecta. Therefore, their dust grain evolution within the ejecta can be compared to the non-magnetic cases but not to our main work presented in this study. In addition, they considered sputtering as a destruction process but neglect any grain-grain collision processes such as fragmentation or vaporization.

To our knowledge, no other work investigates the destruction of dust grains within the clumpy ejecta of a SNR under the influence of a magnetic field. In addition, our model considers fragmentation and vaporization as well as grain sputtering which makes our simulations unique in its kind.

%
\section{Conclusions}
\label{700}
We have studied the dust survival fractions in an over-dense clump during the passage through the reverse shock in Cas~A. Using the MHD code \textsc{\mbox{AstroBEAR}} and the post-processing code \textsc{\mbox{Paperboats}}, the evolution of the gas and dust, respectively, were simulated. Compared to \citetalias{Kirchschlager2019b}, the following improvements and enhancements have been made:

\begin{enumerate}
 \item Spatial resolution: The number of grid cells was increased from 20 to 100 cells per clump radius. The physical resolution per cell amounts to $\Delta_\text{cell}=\unit[10^{14}]{cm}$ ($\sim\unit[6.7]{au}$), which allow us to trace much finer structures in the post-shock gas and dust.
  \item Magnetic fields: The simulations were extended from hydrodynamics to magneto-hydrodynamics. Besides gas and plasma drag, the charged grains are accelerated by magnetic fields and gyrate perpendicular to the magnetic field lines. This leads to a larger number and higher collision velocities of grain-grain and gas-grain collisions. 
  \item Partial vaporization: The collision between two dust grains at high enough energies previously led both grains to be fully vaporized. The new approach allows us to treat the grains individually now. Depending on the collision energy, this allows to partially vaporize both grains, to totally vaporize one of the grains while the other one is partially vaporized, or to totally vaporize both grains. 
 \item Small grain sizes: The range of the initial size distributions is extended to dust grain radii below $10\,$nm. This allows us to trace small grains which are well coupled to the post-shock gas in high-density clumps, experiencing less dust destruction.     
\end{enumerate}

Our SNR ejecta model represents the cloud-crushing scenario in which a planar shock wave is driven into an over-dense clump of gas which is embedded in a low-density gaseous medium. The high-energy shock ($v_\text{shock}=1600\,$km$\,$s${}^{-1}$) can significantly destroy dust grains that formed in the clumps. In order to improve our understanding of the dust destruction, in our simulations we varied the clump densities (gas density contrast $\chi$ between clump and ambient medium), initial magnetic field strengths $B_0$, and initial grain sizes (grain size $a_\text{peak}$ at which the lognormal grain size distribution has its maximum). We summarize for the dust survival fraction $\eta$:

\begin{description}
  \item[{\bf Density contrast:}] We can roughly distinguish between low- ($\chi \sim 50-100$) and high-density contrasts ($\chi>100$). For low-density contrasts, small dust grains (${\sim}1\,$nm) are completely destroyed by sputtering ($\eta\sim0$~per~cent) while large grains (${\sim}1000\,$nm) mostly survive ($\eta\sim100$~per~cent). The transition is gradual. For high-density contrasts,  small grains (${\lesssim}10\,$nm) show significant dust survival fractions (up to $\eta\sim100$~per~cent for $\chi=1000$). Due to the coupling between gas and dust, these grains are not exposed to the hot gas of the ambient medium and are thus protected in the compressed fragments of the shocked clump. Medium sized grains (${\sim}10\,-$ a few $100\,$nm) decouple from the gas and suffer significant dust destruction by sputtering and grain-grain collisions. For larger grains, a substantial fraction of the dust survives ($30-70\,$per~cent for ${\sim}1000\,$nm grains).        
  \item[{\bf Magnetic field strength:}] A magnetic field with an orientation perpendicular to the shock direction causes an additional acceleration of the charged dust grains. The dust grains gyrate around the magnetic field lines which increases the relative velocities between dust grains and gas as well as between grains of different sizes. The more frequent gas-grain collisions enhance the destruction by sputtering. The large relative velocities, on the other hand, affect grain-grain collisions and increase the destruction primarily of large grains. In total, magnetic fields cause a larger destruction compared to the absence of magnetic fields.\newline
  On the other hand, a magnetic field with an orientation parallel to the shock direction has nearly no impact on the dust destruction rate of grains below $10\,$nm while larger grains (${>}10\,$nm) are slightly more easily destroyed; but still at a lower level compared to a perpendicular field.
  \item[{\bf Initial grain sizes:}] As already outlined in \citetalias{Kirchschlager2019b}, the dust survival fractions strongly depend on the grain sizes initially present in the clump. For a density contrast of $\chi=300$, for example, $56\,$per~cent of the dust mass of $1000\,$nm grains can survive, while the survival fraction is only $4\,$per~cent for $100\,$nm grains ($B_0=0$). Therefore, a specific knowledge of the existing grain sizes is mandatory for the analysis of the dust destruction fractions in SNRs.
  \end{description}
 
 These three parameters have a major impact on the dust survival efficiency in the ejecta of SNRs. Moreover, we also varied the dust material (carbon) and the clump size (doubling the clump radius). Though some differences in the survival rates are visible, the impact of the dust material and the clump size is weaker than the impact of the density contrast, magnetic field strength, or the initial grain sizes. 
 
 We are not aware of any other study that investigates the destruction of dust grains within the clumpy ejecta of a SNR under the influence of a magnetic field on the basis of a highly resolved MHD simulation. Considering non-thermal and thermal sputtering, fragmentation, vaporization, coagulation, gas accretion, ion trapping, grain charging, collisional and plasma drag, and Lorentz acceleration, our model includes a multitude of dust processes unique in their number.
 
In a future study we will implement dust destruction by tensile stress induced by suprathermal rotation or charging (Coulomb explosions). Grains can be disrupted into fragments when centrifugal stress from stochastic gas-grain collisions or the electrostatic stress of a charged grain exceed the grains tensile strength (\citealt{Draine1979, Hoang2020}). We also plan to track the depletion of gas-phase species when accounting for grain growth processes such as gas atom accretion and ion-trapping, as well as tracking increases in gas-phase species abundances caused by grain destruction processes. 
Finally, upcoming Cas~A observations of the James Webb Space Telescope have the potential to give us insights into the shape and structure of the ejecta clumps and we will be able to model more realistic clumps of Cas~A. 
  
%
 \section*{Acknowledgements}
FK, FDS and MJB were supported by European Research Council Grant SNDUST ERC-2015-AdG-694520. FK, IDL and NSS have received funding from the European Research Council (ERC) under the European Union’s Horizon 2020 research and innovation programme DustOrigin (ERC-2019-StG-851622). NSS acknowledges the support from the Flemish Fund for Scientific Research (FWO-Vlaanderen) in the form of a postdoctoral fellowship (1290123N). Simulations were performed using the data intensive {\textsc{Peta4-Skylake}} and {\textsc{Peta4-Icelake}} service at Cambridge, supported through DiRAC project ACSP190 (SNDUST) using the Cambridge Service for Data Driven Discovery (CSD3), part of which is operated by the University of Cambridge Research Computing on behalf of the STFC DiRAC HPC Facility (\href{www.dirac.ac.uk}{www.dirac.ac.uk}). The DiRAC component of CSD3 was funded by BEIS capital funding via STFC capital grants ST/P002307/1 and ST/R002452/1 and STFC operations grant ST/R00689X/1. DiRAC is part of the U.K. National \mbox{e-Infrastructure.}
 

%
 \section*{Data availability}
 The data underlying this article will be shared upon reasonable request to the corresponding author.

{\footnotesize
  \bibliography{Literature}
}

 
   \appendix

  \section{Rankine-Hugoniot jump conditions} 
 \label{app_RHJC}
We outline here the Rankine-Hugoniot jump (RHJ) conditions for the MHD case which define the gas conditions of the post-shock region.
 
For the sake of simplicity, we consider the problem in the rest frame of the shock front. Here, the gas velocity $\mathbf{u}$ and the magnetic field $\mathbf{B}$ can be split into components parallel and perpendicular to the shock front direction, $u_\textrm{x}$ and $u_\textrm{p}$ with $u^2_\textrm{x} + u^2_\textrm{p} = |\mathbf{u}|^2$, as well as $B_\textrm{x}$ and $B_\textrm{p}$ with $B^2_\textrm{x} + B^2_\textrm{p} = |\mathbf{B}|^2$, respectively. In the following, the subscripts ``1'' and ``2'' denote the pre-shock and post-shock values. The pre-shock values for the gas density $\rho_1$, gas pressure $P_1$, gas velocity parallel and perpendicular to the shock front direction, $u_\textrm{x,1}$ and $u_\textrm{p,1}$, and parallel and perpendicular component of the magnetic field, $B_{\textrm{x},1}$ and $B_{\textrm{p},1}$, are given. The post-shock values of the gas are then calculated using the RHJ conditions (\citealt{Ryden2009}),
\begin{align}
\rho_i \, u_{\textrm{x},i} &= \text{const}_\text{I}, \label{eq1}\\
\rho_i \, u_{\textrm{x},i}^2 + \frac{1}{8\pi}B^2_{\textrm{p},i} + P_i & = \text{const}_\text{II},\\
\rho_i \, u_{\textrm{x},i}\, u_{\textrm{p},i} - \frac{1}{4\pi} B_{\textrm{p},i}\,B_{\textrm{x},i} &= \text{const}_\text{III},\\
\rho_i \, u_{\textrm{x},i} \bigg(\frac{\gamma}{\gamma-1}\frac{P_i}{\rho_i}  + \frac{1}{2}\left( u_{\textrm{x},i}^2 + u_{\textrm{p},i}^2\right) \hspace*{1.2cm} 
&\nonumber\\
- \frac{1}{4\pi} B_{\textrm{p},i}\left( B_{\textrm{x},i}\, u_{\textrm{p},i} - B_{\textrm{p},i}\, u_{\textrm{x},i} \right)\bigg)&= \text{const}_\text{IV},\\
B_{\textrm{x},i}\, u_{\textrm{p},i} - B_{\textrm{p},i}\, u_{\textrm{x},i} & = \text{const}_\text{V},\\
B_{\textrm{x},i} & = \text{const}_\text{VI} \label{eq6},
\end{align}
where $i\in\left\lbrace 1,2\right\rbrace$ and const${}_{\text{I}...\text{VI}}\in\mathbb{R}$ are constants.

The angles between the magnetic field $\mathbf{B}$ and the shock propagation direction $\mathbf{x}$ are $\Psi_\text{1}$ and $\Psi_\text{2}$, and for a strong shock is
$\Psi_\text{2}=\tan{{}^{-1}(4\,\tan{(\Psi_\text{1})})}$ (\citealt{Draine1979}). The magnetic field of the post-shock medium is then (even for a random orientation of $\mathbf{B}$  in the pre-shock medium) with a high probability close to an orthogonal alignment regarding the shock propagation direction. Cooling and compression of the post-shock gas  further increases this probability. It is therefore justifiable to assume, w.l.o.g, that the magnetic field in the post-shock medium is perpendicular to the shock propagation direction, $\mathbf{B}\perp\mathbf{x}$.

Combining equations~(\ref{eq1}-\ref{eq6}) and the condition $\mathbf{B}\perp\mathbf{x}$, we solve the equation system and obtain for the post-shock quantities after transforming into the observer frame \mbox{($\mathbf{v}=\mathbf{u}+\mathbf{v}_\textrm{shock}$)}
\begin{align}
 v_{\textrm{x},2} & =\frac{3}{4}v_\textrm{shock}-\frac{3}{16\pi}\frac{B_{\textrm{x},1}^2}{\rho_1\,v_\textrm{shock}} ,\\
 v_{\textrm{p},2} & = v_{\textrm{p},1}-\frac{3}{4\pi}\frac{B_{\textrm{x},1}\,B_{\textrm{p},1}}{\rho_1\,v_\textrm{shock}},\\
 \rho_2		  & = 4\rho_1 \left(1 + \frac{3}{4\pi}\frac{B_{\textrm{x},1}^2}{\rho_1\,v_\textrm{shock}^2}\right)^{-1},\\
 P_2   		  & = P_1 + \frac{3}{4}\rho_1\,v_\textrm{shock}^2 - \frac{3}{16\pi}\left(10B_{\textrm{p},1}^2 + B_{\textrm{x},1}^2 \right),\\
 B_{\textrm{x},2} & = B_{\textrm{x},1},\\
 B_{\textrm{p},2} & = 4B_{\textrm{p},1}.
 \end{align}
These post-shock conditions have been implemented in \mbox{AstroBEAR} and are solved by it automatically. Please note, that for $B_\textrm{x} = B_\textrm{p} =0$ the RHJ conditions of the MHD case transform to the RHJ conditions of the hydrodynamical case.


\label{lastpage}


\end{document}